\documentclass[aps,prl,preprint,superscriptaddress]{revtex4-1}

\usepackage{makeidx}
\usepackage[verbose]{placeins}
\usepackage{multirow}
\usepackage{graphicx}
\usepackage{amsmath}
\usepackage{amsfonts}
\usepackage{epstopdf}
\usepackage{textcomp}
\usepackage{epstopdf}
\usepackage{tabularx}
\usepackage{ulem} 
\usepackage{color}

\usepackage{tabularx}
\usepackage{ulem} 
\usepackage{color}

\usepackage{natbib}
\usepackage{bm}%
\usepackage[colorlinks=true,linkcolor=blue]{hyperref}%
\expandafter\ifx\csname package@font\endcsname\relax\else
 \expandafter\expandafter
 \expandafter\usepackage
 \expandafter\expandafter
 \expandafter{\csname package@font\endcsname}%
\fi
\makeindex

\begin{document}

\title{Modification of spintronic terahertz emitter performance through defect engineering}

\author{Dennis M.~Nenno}
\author{Laura Scheuer}

\affiliation{Fachbereich Physik and Landesforschungszentrum OPTIMAS, Technische Universit\"{a}t Kaiserslautern, Erwin-Schr\"{o}dinger-Str.~56, 67663 Kaiserslautern, Germany}

\author{Dominik Sokoluk}
\affiliation{Fachbereich Elektro-Informationstechnik and Landesforschungszentrum OPTIMAS, Technische Universit\"{a}t Kaiserslautern, Erwin-Schr\"{o}dinger-Str.~11, 67663 Kaiserslautern, Germany} 

\author{Sascha Keller}
\affiliation{Fachbereich Physik and Landesforschungszentrum OPTIMAS, Technische Universit\"{a}t Kaiserslautern, Erwin-Schr\"{o}dinger-Str.~56, 67663 Kaiserslautern, Germany}

\author{Garik Torosyan}
\affiliation{Photonic Center Kaiserslautern, Kaiserslautern 67663, Germany}

\author{Alexander Brodyanski}
\author{J\"{o}rg L\"{o}sch}
\affiliation{Institut f\"{u}r Oberfl\"{a}chen- und Schichtanalytik (IFOS) and Landesforschungszentrum OPTIMAS, Trippstadter Str.~120, 67663 Kaiserslautern, Germany} 

\author{Marco Battiato}
\affiliation{Division of Physics and applied physics, Nanyang Technological University, Singapore}

\author{Marco Rahm}
\affiliation{Fachbereich Elektro-Informationstechnik and Landesforschungszentrum OPTIMAS, Technische Universit\"{a}t Kaiserslautern, Erwin-Schr\"{o}dinger-Str.~11, 67663 Kaiserslautern, Germany} 

\author{Rolf H. Binder}
\affiliation{College of Optical Sciences, University of Arizona, Tucson, AZ 85721, USA}
\author{Hans C.~Schneider}
\author{Ren\'{e} Beigang}
\affiliation{Fachbereich Physik and Landesforschungszentrum OPTIMAS, Technische Universit\"{a}t Kaiserslautern, Erwin-Schr\"{o}dinger-Str.~56, 67663 Kaiserslautern, Germany}

\author{Evangelos Th.~Papaioannou\email{papaio@rhrk.uni-kl.de}}
\thanks{Author to whom correspondence should be addressed. Email: papaio@rhrk.uni-kl.de}

\affiliation{Fachbereich Physik and Landesforschungszentrum OPTIMAS, Technische Universit\"{a}t Kaiserslautern, Erwin-Schr\"{o}dinger-Str.~56, 67663 Kaiserslautern, Germany}

%


\begin{abstract}

Spintronic ferromagnetic/non-magnetic heterostructures are novel sources for the generation of THz radiation  based on spin-to-charge conversion in the layers. The key technological and scientific challenge of THz spintronic emitters is to increase their intensity and frequency bandwidth. Our work reveals the factors to engineer spintronic Terahertz generation by introducing the scattering lifetime and the interface transmission for spin polarized, non-equilibrium electrons. We clarify the influence of the electron-defect scattering lifetime on the spectral shape and the interface transmission on the THz amplitude, and how this is linked to structural defects of bilayer emitters. The results of our study define a roadmap of the properties of emitted as well as detected THz-pulse shapes and spectra that is essential for future applications of metallic spintronic THz emitters.

\end{abstract}

\flushbottom
\maketitle

\thispagestyle{empty}

\section*{Introduction}
Recent studies in spintronics have highlighted ultrathin magnetic metallic multilayers as a novel and promising class of broadband terahertz radiation sources~\cite{Walowski:2016bt}. Such spintronic heterostructures consist of ferromagnetic (FM) and non-magnetic (NM) thin films. When triggered by ultrafast femtosecond (fs) laser pulses, they generate pulsed terahertz (THz) electromagnetic radiation due to the inverse spin Hall effect (ISHE), a mechanism that converts the spin currents originating in the magnetized FM layer into transient transverse charge currents in the NM layer resulting in THz emission~\cite{Kampfrath2013}. Different strategies have been followed in order to explore the THz amplitude and bandwidth of the signal: different material compositions of FM/NM systems with a variety of thicknesses~\cite{Torosyan2018,Seifert2016,ADMA:ADMA201603031,ADOM:ADOM201600270,Papaioannou2018,Qiu:18}, ferri- and antiferromagnetic metal/Pt structures~\cite{spin2017,Albrecht2018}, spintronic emitters assisted by metal-dielectric photonic crystal~\cite{Haifeng2018}, metallic trilayer structures with different interface materials~\cite{Li_2018,Seifert_2018,Li_2019}, THz emission from Rashba type interfaces~\cite{PhysRevLett.121.086801,PhysRevLett.120.207207} and THz emission using different excitation wavelengths~\cite{Papaioannou2018,Herapath2019}. The spin-to-charge-conversion mechanism was additionally probed in metallic and insulating magnetic/NM interfaces~\cite{Ilya2017,doi:10.1021/acs.nanolett.7b04538,yigPt2018} showing, however, much lower efficiency of the THz emission compared to the metallic magnetic layers.
Even though the transfer of a spin current from a FM to a NM layer (that is the source of THz emission) is a highly interface-sensitive effect, no correlation between the structural quality of the interface and crystal properties of the metallic components on the signal strength and spectrum has been established. The few existing papers have hinted at contradicting results. A direct comparison of an epitaxial Fe(3\,nm)/Pt(3\,nm) bilayer with a signal-optimized polycrystalline CoFeB/Pt structure with the same layer thicknesses revealed a comparable THz signal strength~\cite{Seifert2016}. Contrariwise, a significant increase in signal amplitude between Fe/Pt emitters grown epitaxially on MgO (100) substrates compared to polycrystalline emitters grown on sapphire substrates was reported in Ref.~\citenum{Torosyan2018}, but not further investigated. Similarly, the better crystal quality of a CoFeB layer, controlled by the annealing temperature, has significantly enhanced THz emission intensity~\cite{Sasaki}. Besides the numerous studies, the understanding and the structural engineering of THz intensity and spectral bandwidth of spintronic emitters remains an uncharted territory. 

In this work, we reveal the significance of the sample growth, i.e., the defect density and interface quality, for the enhancement of the emitted THz signal amplitude and bandwidth. We address the influence of single-crystallinity and the local FM/NM interface morphology on the THz emission. We link the structural properties of the metallic bilayers with the THz signal amplitudes and spectra of the emitters. We compare our experimental findings with a theoretical model based on the Boltzmann transport equation that accounts for the differences in elastic electron scattering lifetime in the layers and for the transmission of spin-polarized hot carriers at the Fe/Pt interface. 

\section*{Results}

\begin{figure}
	\centering
	\includegraphics[width =1.0 \columnwidth]{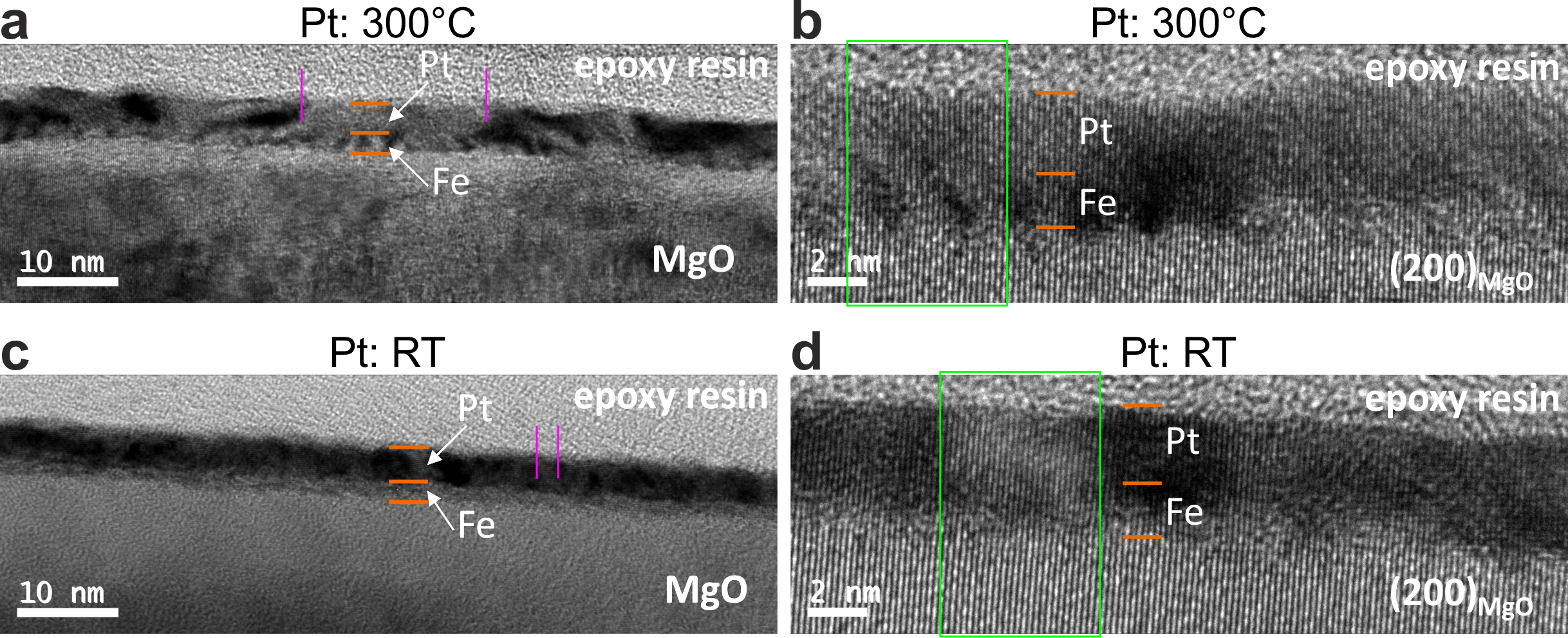}
	\caption{\label{fig:TEM0} Cross-sectional zero-loss energy filtered transmission electron microscopy images for two magnifications of the Fe\,(2\,nm)\,/\,Pt\,(3\,nm) bilayers on MgO\,(100), where the Fe layers are grown at 300\,$^\circ$C, but the Pt layers are grown at 300\,$^\circ$C (\textbf{a}\, and \,\textbf{b}) and room temperature (RT, \textbf{c}\, and \,\textbf{d}), respectively. The dark-contrast regions correspond to the local nm-sized areas, being twisted relative to the MgO substrate. This twisting (small rotations or tilts) causes strong diffraction contrast. In \textbf{a}\, and \,\textbf{b}, such twisted regions are noticeably fewer and more localized. This produces relatively large and weakly deformed (almost deformation-free) lateral regions (marked with vertical lines). In \textbf{c}\, and \,\textbf{d}, the opposite situation occurs: the weakly deformed areas are laterally small and the strongly deformed regions dominate the bilayer within both individual layers. The high contrast between two individual layers in the RT-Pt sample is qualitatively similar to the one in the case of amorphous (or ideal polycrystalline) layers, where it is caused by a big difference in the atomic form-factor between Fe and Pt. It is a direct evidence for a deformed crystal growth of individual layers. The relatively weak contrast difference between the Fe layer and the single crystal MgO points to a good crystallographic adaption of the Fe layer to the substrate.  Together with the previous statement, this implies that the Pt layer is the most laterally deformed one. On the other side, the loss of contrast difference between the Fe layer and Pt for the 300\,$^\circ$C-Pt sample reveals good epitaxial growth of the complete bilayer.}
	
\end{figure}

\begin{figure}
	\centering
	\includegraphics[width =0.5 \columnwidth]{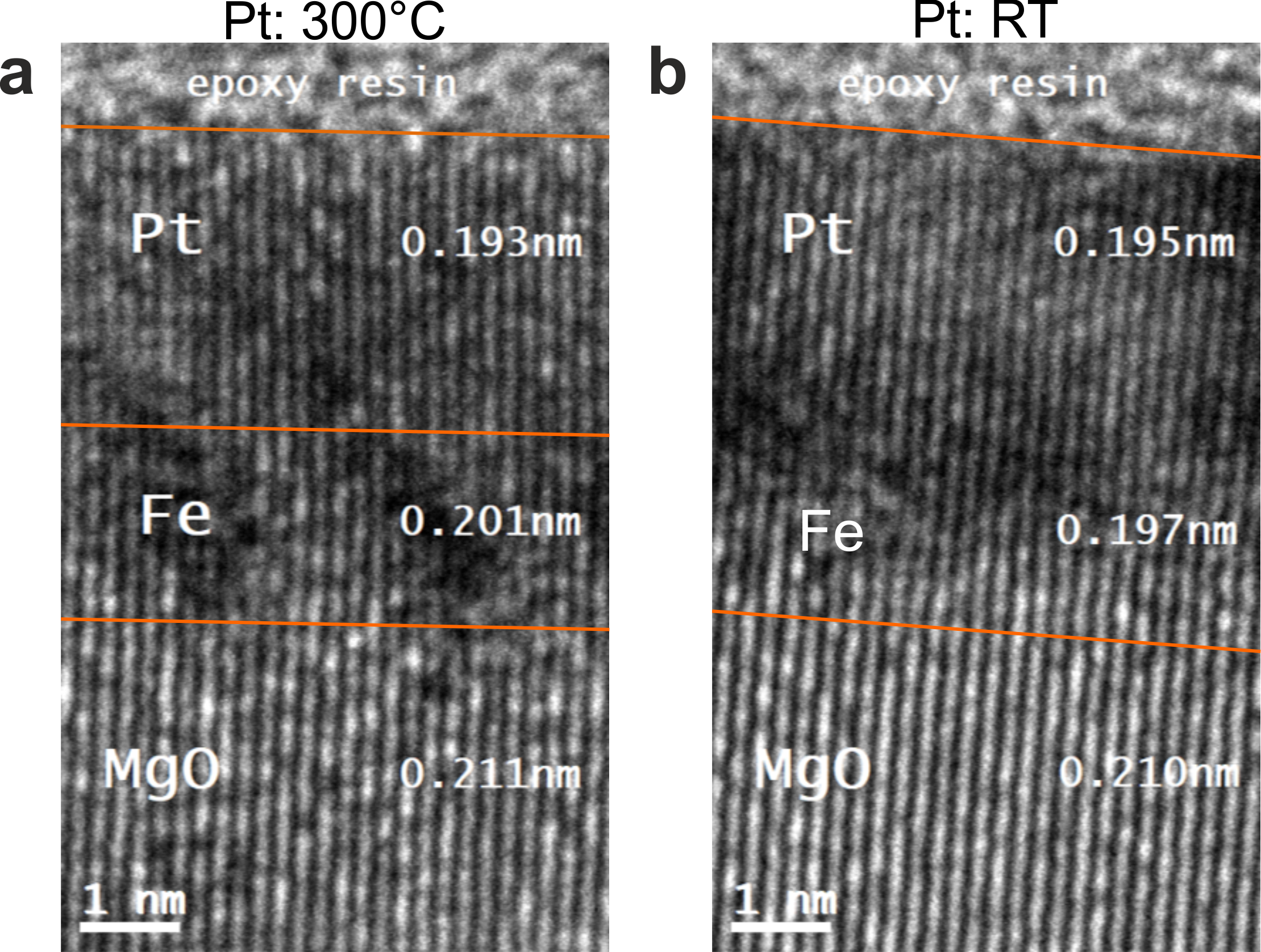}
	\caption{\label{fig:TEM} A magnification of the selected regions of Fig.~\ref{fig:TEM0}\,\textbf{b} and \textbf{d}, of the Fe\,(2\,nm)\,/\,Pt\,(3\,nm) bilayers on MgO\,(100): \textbf{a} refers to the 300\,$^\circ$C-Pt and \textbf{b} to the RT-Pt sample, respectively. The atomic planes within the Fe/Pt system are resolved. The MgO crystal lattice is slightly tilted relative to the $z$-axis of the image frame (1.1\,$^\circ$ and 4.3\,$^\circ$ in \textbf{a} and \textbf{b} respectively). The measurements of the lattice spacing within the Pt and Fe layers, shown in \textbf{a}, reveal the values that coincide with the ones for stress-free single crystals. 
	Strong modification of the local lattice spacing and subsequently the quality of the interface between Fe and Pt for the RT-Pt sample is shown in \textbf{b}. Here, the Fe layer is characterized by the (110) lattice spacing of about 0.197\,nm that is substantially smaller than expected for stress-free single-crystal Fe $d_\text{Fe(110)}$\,=\,0.2027\,nm and lies very close to the (200) lattice spacing of Pt (measured value of 0.195\, nm).}
\end{figure}

\subsection*{Epitaxial and polycrystalline growth} In order to address the generation of THz radiation by spintronic emitters with varying structural parameters, we investigate fully epitaxial stress-free, deformed epitaxial and non-epitaxial bilayers. We use Fe/Pt bilayer samples as a model system with different interfaces and crystal quality. We focus on Fe\,(2\,nm)\,/\,Pt\,(3\,nm) bilayers since these thicknesses are optimized to provide the highest THz amplitude~\cite{Torosyan2018,Papaioannou2018}. The Fe/Pt samples were either grown epitaxially on MgO\,(100) or polycrystalline on sapphire Al$_2$O$_3$\,(0001) substrates~\cite{Evangelos,Andres,Keller2018}.
Conventional and high resolution zero-loss energy filtered transmission electron microscopy (HR EFTEM) measurements shown in Fig.~\ref{fig:TEM0}, present two cases: one Fe(2\,nm)~/Pt(3\,nm) sample grown entirely at 300\,$^\circ$C (300\,$^\circ$C-Pt-sample) and another Fe\,(2\,nm)\,/\,Pt\,(3\,nm) grown at 300\,$^\circ$C (Fe layer) and at room temperature (RT) (Pt layer), (RT-Pt-sample). 
In all images of Fig.~\ref{fig:TEM0}, the thin-film system is clearly distinguishable from the substrate. 
At higher magnifications, the epitaxially grown atomic planes within the Fe/Pt system are well resolved. A contrast within the thin-film system varies both in longitudinal and transversal directions relative to the substrate that points toward different local crystallographic orientations within the thin-film systems. An absence of detectable grain boundaries within the bilayer indicates that local disorientation becomes possible with the help of local defects (dislocations, vacancies) or/and elastic deformations, which is possible within very thin layers. The comparison of the two different samples shows that the deformation in the RT-sample is substantially stronger than in the one grown at 300\,$^\circ$C. The deformation can be understood as local nm-sized areas that are twisted (small rotations or tilts) relative to the MgO substrate. These twisted regions are noticeably fewer for the 300\,$^\circ$C-Pt-sample, which results in relatively large, almost deformation-free lateral regions. The RT-Pt-sample has strongly deformed regions that dominate the bilayer within both individual layers. These microscopic findings reflect the quality of the interface: the 300\,$^\circ$C-Pt-sample possesses an undisturbed Fe/Pt interface for lateral sizes of about 20\,nm (see vertical lines in Fig.~\ref{fig:TEM0},\textbf{a}). The corresponding defect-free lateral area for the more strongly deformed crystal for the RT-Pt sample reaches values close to 3\,-\,4\,nm (see vertical lines in Fig.~\ref{fig:TEM0},\textbf{c}). 
For even larger magnifications, Fig.~\ref{fig:TEM}\,\textbf{a} and \textbf{b}, the cause of the differences in deformations is visible: Fig.~\ref{fig:TEM}\,\textbf{a} for 300\,$^\circ$C-Pt sample reveals the bilayer and epitaxial character of the thin-film system that remains clearly discernible. The lattice spacing within the regions of the Pt and Fe layers provide lattice constants for practically stress-free single crystals. On the other hand, in Fig.~\ref{fig:TEM}\,\textbf{b} (RT-Pt-sample) the Fe layer is characterized by a (110) lattice spacing of about 0.197\,nm that is substantially smaller than strain-free Fe and lies very close to the (200) lattice spacing of Pt. The relaxation processes of the strained lattices lead to local crystallographic disorientations in both cases. However, the degree of disorientation is noticeably stronger and disorientations occur more frequently in the RT-Pt-sample. It is worth noticing that the presented images of the HR EFTEM investigations are representative for the entire layers of interest. The differences in the local microstructure have dramatic influence on the ability of the samples to emit THz radiation.

\begin{figure*}
	\centering
	\includegraphics[width =0.9 \columnwidth]{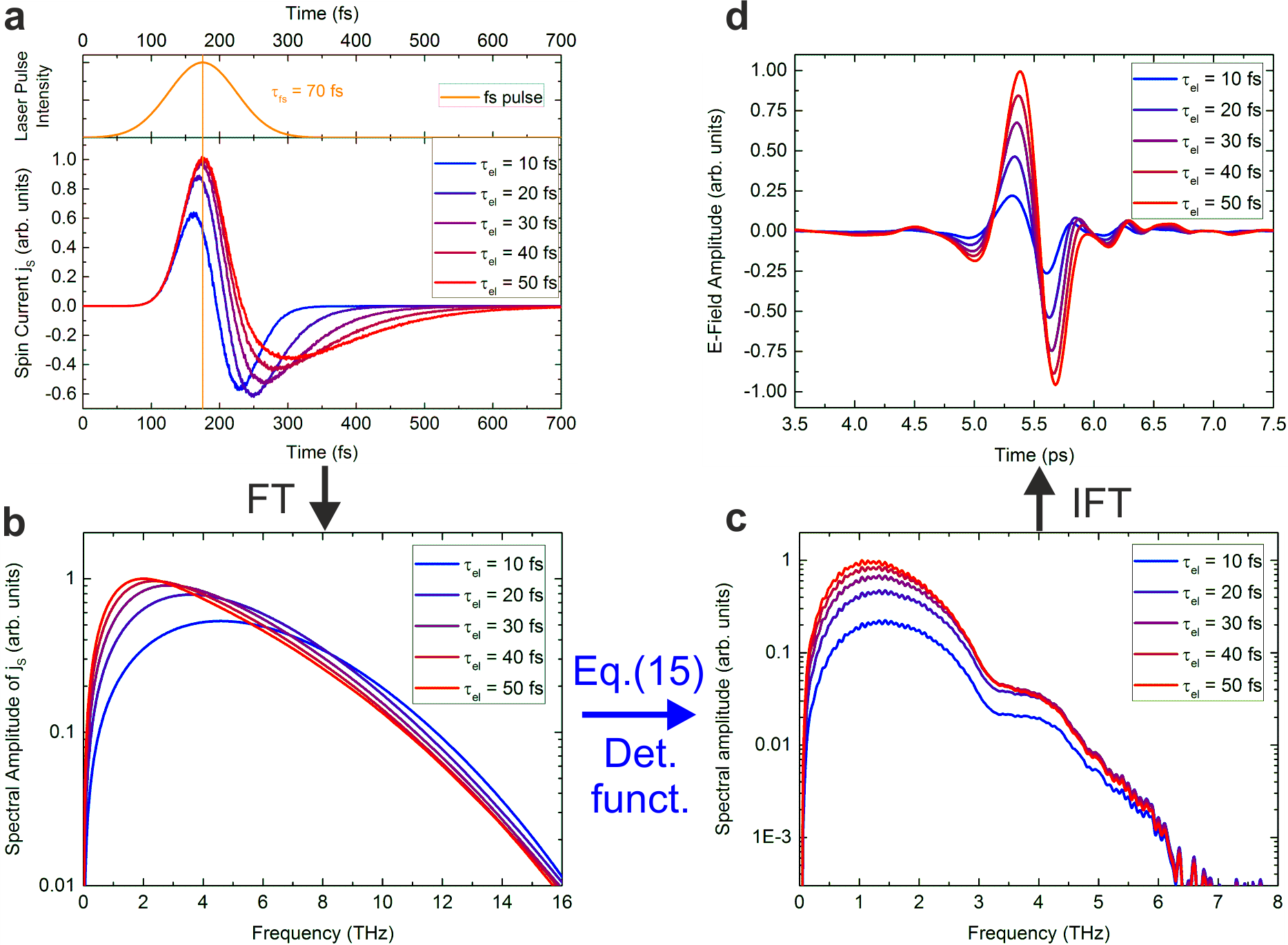}
	\caption{\label{fig:003} Results of the numerical simulation of the Boltzmann transport equation and transfer-matrix formalism for the spintronic Fe\,(2\,nm)\,/\,Pt\,(3\,nm) emitter for different elastic scattering lifetimes $\tau_\text{Pt}$ of the electrons including scattering with defects inside the Pt layer (all plots normalized to the maximum). \textbf{a} (top graph) Temporal evolution of the laser pulse intensity with a pulse width of 70\,fs and its maximum intensity at $t_\text{max}$\,=\,175\,fs. (bottom graph) Simulated spin current $j_\text{s}$ inside the Pt layer (averaged over the Pt thickness). The spin current is proportional to the charge current $j_\text{c}$ through the ISHE (Eq.~\ref{eq:spinhall}). \textbf{b} Spectral amplitudes of the spin current (and consequently of $j_\text{c}$) inside the Pt layer given by the Fourier transform (FT) of \textbf{a} (phase of the components not shown here). \textbf{c} Simulated spectral amplitude of the THz-E-field measured after propagation through 500\,$\mu$m of MgO and accounting for the detector function. This was calculated using \textbf{b}, the transfer matrices from Eq.~\ref{eq:M} (phase of the components experience dispersion considered in the transfer matrices) and the detector response function (see Fig.~\ref{fig:resp}). \textbf{d} E-field amplitude of the simulated THz-wave measured by the detector as given by the inverse Fourier transformation (IFT) of \textbf{c}. Due to dispersion, the low-frequency components arrive earlier than the high-frequency components (the phase information of the components was re-used for the IFT). Note the change of units from \textbf{a} to \textbf{d}.}
\end{figure*}

\subsection*{Capturing the structure-dependent THz emission with Boltzmann transport simulations} The electron dynamics that determines the evolution of THz emission from fully disoriented crystallites (maximum number of defects) to fully epitaxial bilayers (minimum number of defects) is theoretically described a model based on the Boltzmann transport equation (BTE) which captures the following scenario~\cite{Hurst:2018bz,Manfredi:2005ba}.
Femtosecond laser pulses excite electrons from below to above the Fermi level. The excited quasi-free carriers ("hot carriers") then move through the structure and scattering processes are responsible for the eventual return to equilibrium. The BTE (see methods, Eq.~\ref{eq:bte2}) determines the space- and time-dependent carrier distribution function of electrons with a spin-orientation~\cite{Nenno2018PIC}. The absorption of a fs laser pulse inside the metallic layers is governed by typical optical parameters (see methods) and we follow the idea originally proposed in Ref.~\citenum{Kampfrath2013} to calculate absorption, electron dynamics and THz emission. Our numerical approach for solving the BTE in multilayer systems is presented in Ref.~\citenum{Nenno_arxiv_THz}. As result of our simulation, the time-dependent carrier distribution for hot spin-up and spin-down electrons is obtained for the Pt layer and will be used for calculating the spin current. Critical material parameters concerning the structural features of the heterostructures are the elastic scattering lifetime $\tau_\text{el}$ that correlates with the overall defect density in the Fe and Pt layers and the Fe/Pt interface transmission coefficient $T$ that is influenced by local lattice defects at the interface. Other material parameters, such as carrier velocities at different energies as well as their lifetimes are taken from literature (from Ref.~\citenum{Kaltenborn:2014du} for Fe and from Ref.~\citenum{Zhukov} for Pt).

\subsection*{Influence of the elastic electron-defect scattering lifetime on the THz emission} First, we discuss the variation of the electron-defect scattering lifetime. Typical elastic scattering times are on the order of 50\,fs and can be decreased by increasing the number of defects. In Fig.~\ref{fig:003}\,\textbf{a}, the induced spin current averaged over the Pt layer is shown. The results were calculated for Fe(2nm)~/~Pt(3nm) when irradiated from the Pt side. All signals show a bipolar pulse shape (one maximum and one minimum) with an initial peak close to the maximum of the laser excitation, $t_{\mathrm{max}}=$175\,fs. The signal does not vanish until 500\,fs after its onset. The shape of the time-dependent spin current for small Fe/Pt layer thicknesses on the order of typical mean-free paths (carrier lifetime 10\,fs times carrier velocity 1\,nm/fs) can be explained as follows: reflections of the electrons at the metal/insulator and metal/air interfaces allow the charge carriers to traverse the metal stack many times and the net spin current is then formed as a combination of the temporal evolution of the fs pulse intensity and of the difference of the charge carrier velocities depending on their spin orientation.
For longer elastic scattering lifetimes, the spin-polarized carriers contribute to a longer-lasting and stronger signal. The magnitude of the spin current decreases with smaller lifetimes or, equivalently, larger scattering rates. The minority carrier peak (negative sign) is enhanced for shorter lifetimes due to the earlier onset of majority carrier relaxation in Pt which propagate initially at higher energies. For smaller lifetimes, the spin current decays faster (shorter signal, especially for the negative part) which can be understood as a faster relaxation of the spin system towards equilibrium. 
The spin current spectra shown in Fig.~\ref{fig:003}\,\textbf{b} are the Fourier transforms of the spin currents from Fig.~\ref{fig:003}\,\textbf{a}. Since the magnitude of the spin-polarized charge currents decreases with lower scattering time, the shape of the spin currents becomes more symmetric and the spectrum has a lower maximum value. The position of the maximum shifts towards higher frequencies, as the relaxation happens on a shorter timescale. Another main difference of the spectra is the absence of low-frequency components for shorter electron-defect lifetimes.

The aim of the following calculations is to simulate the THz-signal measured by the photoconductive antenna (PCA) starting from the results of the electron-transport calculation. To obtain the spectral electric field amplitude $E(\omega)$ of the THz wave detected by the PCA, cf.~Fig.~\ref{fig:003}\,\textbf{c}, from the spin current spectrum (Fig.~\ref{fig:003}\,\textbf{b}), multiple steps have to be considered in the following:
(i) The in-plane oscillating charge current $j_\text{c}(\omega)$ in the Pt, which generates the THz emission, is proportional to the oscillating spin current $j_\text{s}(\omega)$ by the ISHE. The proportionality constant is defined as the spin-Hall-angle (see Eq.~\ref{eq:spinhall}). (ii) We analytically solve the Green function for the inhomogeneous Helmholtz equation and use the transfer-matrix formalism to account for the absorption and reflection of the THz wave by the metallic and insulator layers (see methods, Eq.~\ref{eq:helmholtz} and following). (iii) The PCA detector itself modifies the experimentally measured spectra due to its frequency-dependent sensitivity (see methods). 

Figure~\ref{fig:003}\,\textbf{c} then shows the results of these calculations. The larger spectral amplitudes at lower frequencies for smaller scattering lifetimes prevail in the THz-spectra in Fig.~\ref{fig:003}\,\textbf{c}. After the propagation of the THz pulse through the lens, the pulse is emitted into free space and is directed by the THz optics onto the antenna. High-resistivity silicon is known for its small, frequency-independent attenuation of THz radiation. Therefore, the influence of the Si-lens on the spectrum in Fig.~\ref{fig:003}\,\textbf{c} is small. It is outside the scope of the simulation to numerically model the influence of the 6\,mm thick lens. Its effect on the spectrum is included in the extracted detector function (see methods). However, optical transmission, absorption and reflection effects in both the metallic and the MgO layer, as well as at the interfaces, have been taken into account by the transfer-matrix formalism. The general discussion for the spectra concerning $\tau_\text{el}$ remains valid, however their bandwidth is decreased due to the strong optical absorption in MgO occurring above 3\,THz~\cite{RenGuanhua:825001}. 

The scattering lifetime determines the maximum frequency of the spectrum as well as its bandwidth (FWHM). However, the exact position of the maximum frequency is difficult to extract with adequate accuracy using a photoconductive switch as the detector, such that in our discussion, we will use the spectral width for comparison with the experimental data.

To obtain the time-dependent, detected THz-pulses, and in order to compare them with the measured pulses, the spectra from Figure~\ref{fig:003}\,\textbf{c} and their phase information (influenced by the dispersion of the materials involved) are used for an inverse Fourier transformation (IFT). The temporal pulse shape shown in Fig.~\ref{fig:003}\,\textbf{d} is now distorted, since different frequency components excited in the Pt layer propagate through the structure at varying velocities. Lower frequency contributions arrive earlier whereas high frequency components account for an oscillating tail of the signal that decreases to less than 10\,\% of the peak value approx.~2\,ps after the onset of the signal. Compared to the original pulses in \textbf{a}, the THz pulses in \textbf{d} are considerably extended in time due to the absence of the high-frequency components and are very similar to the experimentally measured pulses discussed in detail in the next section in Fig.~\ref{fig:ExpTHz}.

\subsection*{Influence of the Fe/Pt interface transmission on the THz emission} 
\begin{figure}[t]
	\centering
	\includegraphics[width=0.55\textwidth]{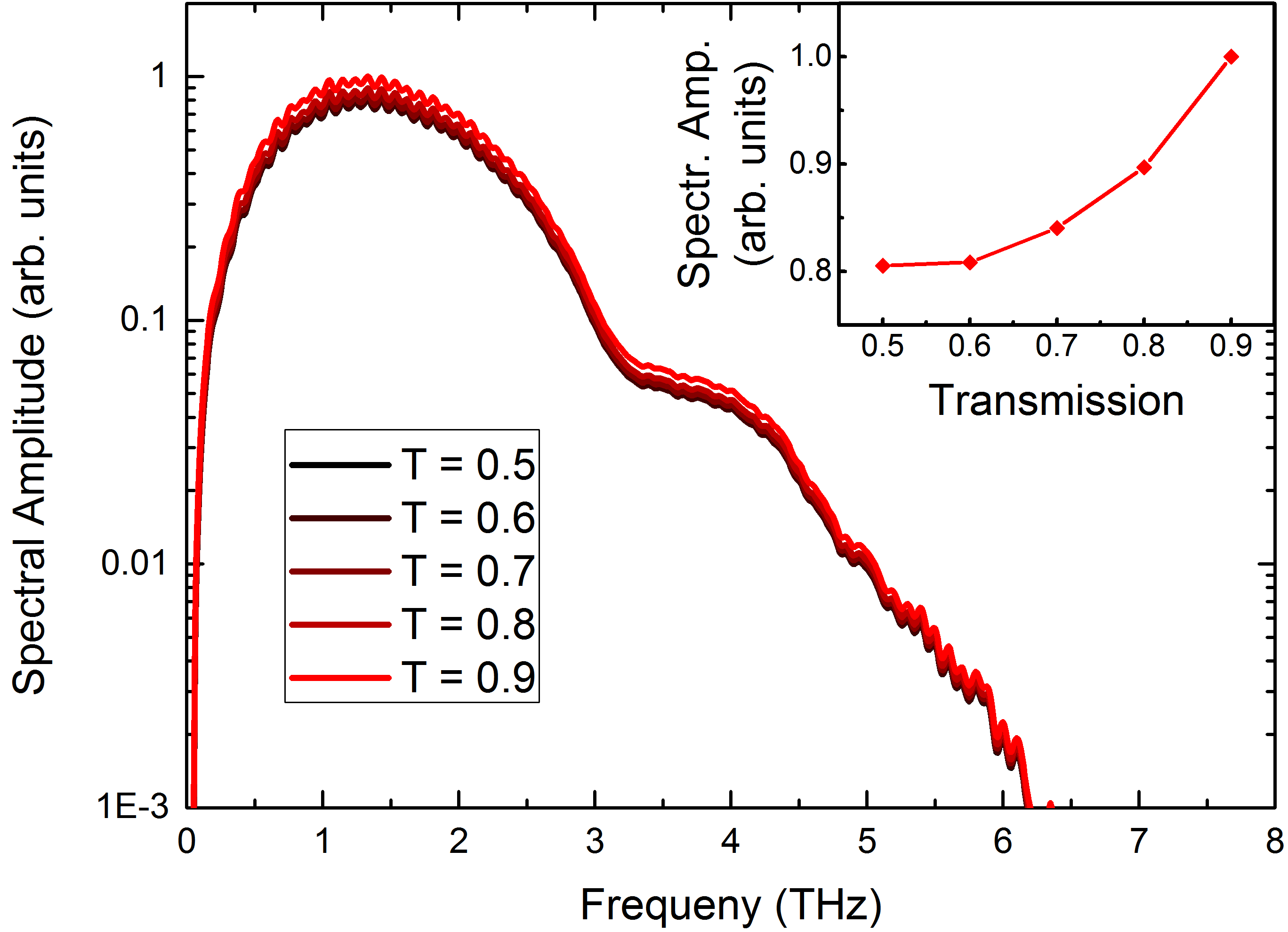}
	\caption{Simulated spectra for the Fe\,(2\,nm)\,/\,Pt\,(3\,nm) emitter on MgO ($\tau_\text{el} =$ 30\,fs) for different interface transmissions $T=T_\text{up}=T_\text{down}$ from 0.5 to 0.9. $T$ refers to the ability of the optically-excited, spin-polarized carriers to cross the interface. Decoherence or dissipation of the spin current at the interface is neglected. The band structures of Fe and Pt are not considered for the transmission of hot carriers between the materials. Inset: Peak amplitudes of the spectra depending on the interface transmission $T$, normalized to the maximum value. \label{Fig:Results4}}
\end{figure}

So far, we have correlated different electron scattering lifetimes (due to changes in the defect density) with the amplitude and the bandwidth of the THz emission. 
In order to fully explain the evolution in the signal magnitude, the Fe/Pt interface transmission $T$ for the hot carriers has to be taken into account. The previous model assumed a perfect transmission of $T$\,=\,1 (no reflection $R$\,=\,0 at the interface) for both majority and minority electrons at all energies above the Fermi level. As depicted by the results of the simulations in Fig.~\ref{Fig:Results4}, the shape of the spectra does not change with reducing the transmission, but the amplitude of the THz signal decreases gradually with lower $T$. As shown in the inset of Fig.~\ref{Fig:Results4}\, the peak amplitude decreases monotonically from $T=0.9$ to $T=0.5$ with a reduction of about 20\,\% in amplitude. Therefore, even a moderate deviation from almost perfect transparency will still be visible in experiments. Although $T$ is an energy- and material-dependent quantity, we will regard it as an independent parameter in our simplified model. This is partly due to the fact that $T(E)$ is very difficult to calculate ab-initio (in particular for strained systems) and partly because it simplifies the physical picture for this study.

\subsection*{Experiment: Terahertz emission from structurally modified heterostructures} 
\begin{figure*}
	\centering
	\includegraphics[width =1.00 \columnwidth]{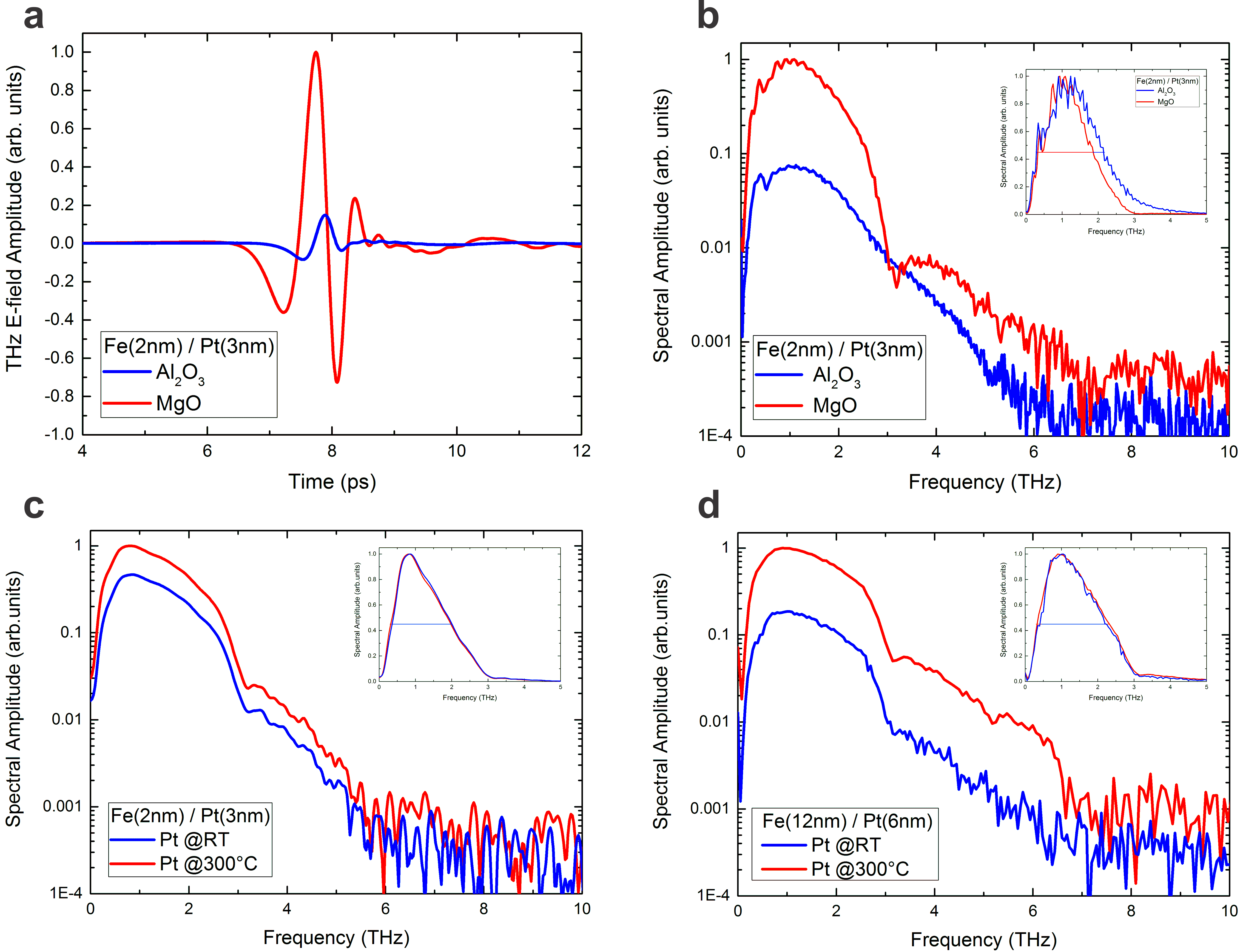}
	\caption{\label{fig:ExpTHz} \textbf{a}  Experimental THz-E-field amplitudes of the Fe\,(2\,nm)\,/\,Pt\,(3\,nm) emitters grown at 300\,$^\circ$C on MgO and Al$_2$O$_3$ in the time domain and \textbf{b}  their corresponding spectra. The emitter grown on MgO exhibits a stronger dispersion compared to the one grown on sapphire as the MgO itself has multiple absorption lines above a frequency of 3\,THz (see methods, Fig.~\ref{Fig:Group_Phase_Velocity_MgO}) while the group and phase velocities of THz waves in Al$_2$O$_3$ are mostly frequency independent up to 10\,THz. \textbf{c}  Spectra of the Fe\,(2\,nm)\,/\,Pt\,(3\,nm) grown on MgO where the Pt layers have been grown at room temperature (RT) or 300\,$^\circ$C, respectively. \textbf{d}  Spectra of two thicker samples, Fe(12\,nm)\,/\,Pt(6\,nm) grown on MgO with different crystal quality of the Pt layer. In both \textbf{c} and \textbf{d} the epitaxial samples exhibit singificantly larger THz-E-field amplitudes. The insets show the spectral width of the corresponding samples. The large change in the amplitude is present in all measurement geometries (detected pulse from the Pt side, with and without the lens), demonstrating the intrinsic origin of the effect.}
\end{figure*}

We now correlate the simulated behavior of the THz-emitters with experiments. The THz pulses and spectra obtained from the spintronic Fe\,(2\,nm)\,/\,Pt\,(3\,nm) emitters grown on MgO and sapphire are shown in Fig.~\ref{fig:ExpTHz}\,\textbf{a,~b}): the E-field peak-to-peak amplitude of the emitter on sapphire is about 90\,\% smaller than the one on MgO. The THz pulse of the emitter on sapphire is slightly compressed in time, which indicates a larger contribution of high-frequency components or equivalently lower dispersion (which is also true for sapphire~\cite{Querry1985}). 

The epitaxial emitter grown on MgO shows a large signal enhancement at frequencies lower than 3\,THz. In the frequency range between 3 and 6\,THz, both emitters have a similar spectral amplitude. The epitaxial growth enhances the signal strength of the lower frequencies by about one order of magnitude. The large change in the amplitude is also present when the pulse is detected from the Pt-side, and thus is independent of the use of a Si-lens, cf.~Fig.~\ref{fig:PtsideMgO}.

\begin{figure}[t]
	\centering
	\includegraphics[width =1.0 \columnwidth]{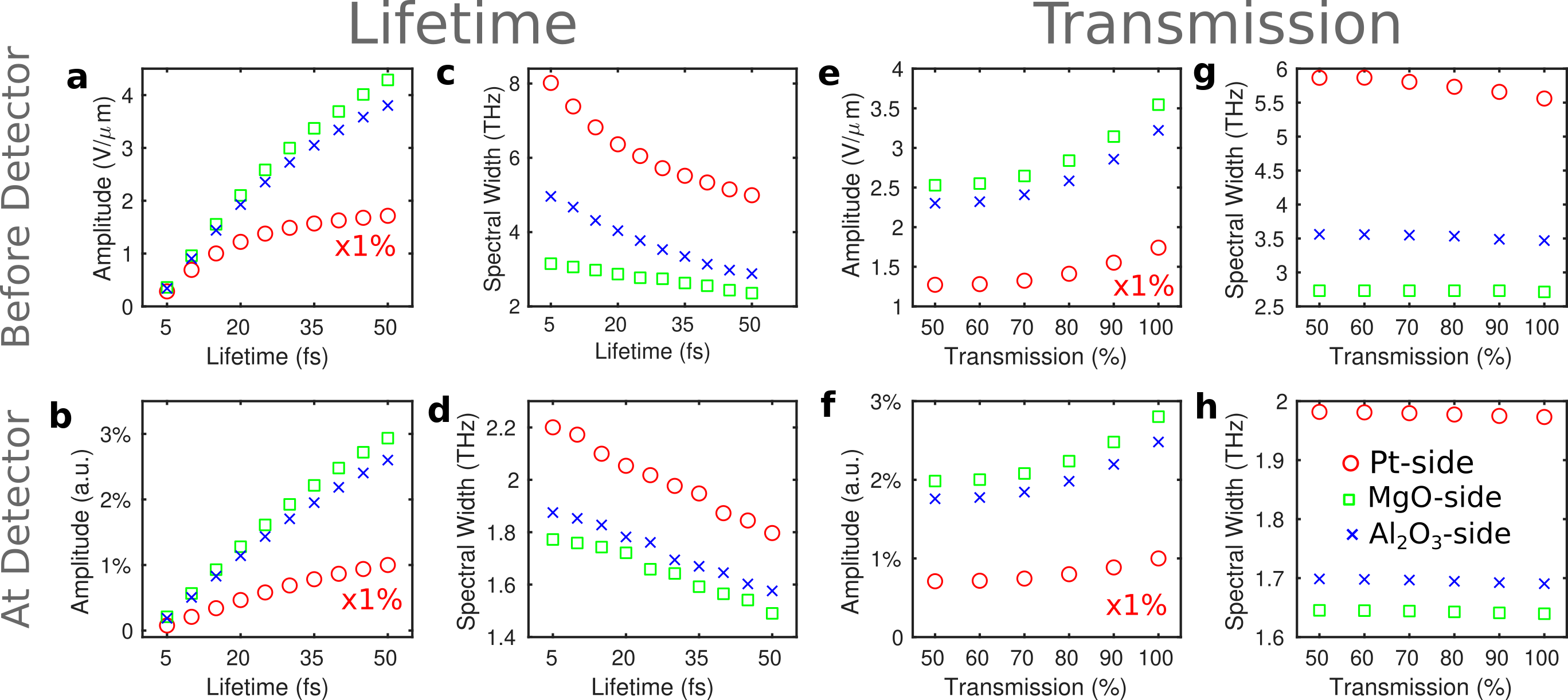}
	\caption{\label{fig:tau-T} Calculated amplitude (a,\,e) and spectral width at half maximum (c,\,g) of the emitted THz pulse for different scattering lifetimes $\tau_\text{el}$ (a,\,c), and for different transmission coefficients $T$ (e,\,g) for the case of $\tau_\text{el}=30 \mathrm{fs}$. Three different measurement scenarios are simulated: the pulse is detected directly from the Pt side (circles), through the MgO substrate (rectangles) and through the Al$_2$O$_3$ substrate (crosses). After taking into account the detector response, the amplitudes (b,\,f) and spectral widths (d,\,h) change. The spectral response of our detection scheme largely influences the amplitude of the THz emission and diminishes the ability to detect shifts of the spectra. Pt values have been multiplied by 0{.}01 (1\,\%) for clarity.}
\end{figure}

Next, we discuss the spectra of the Fe\,(2\,nm)\,/\,Pt\,(3\,nm) emitters grown on MgO shown in Fig.~\ref{fig:TEM0} and Fig.~\ref{fig:TEM}.
When looking at the spectral amplitude of the THz pulses in Fig.~\ref{fig:ExpTHz}\,\,\textbf{c}) for fully epitaxial and partially epitaxial samples, it is apparent that the spectra possess similar shapes: all curves exhibit their maximum signal at around 850\,GHz and have the same characteristic decrease in amplitude at around 3\,THz because of the absorption profile of MgO~\cite{Torosyan2018}. In addition, the curves are shifted in amplitude (log-scale) by a factor of about 2. The fully epitaxial bilayer grown at 300\,$^\circ$C possesses the larger signal. 

Both samples maintain the spectral shape of the THz radiation despite the different degrees of deviation from perfect epitaxy. Here, the interface transmission parameter, Fig.~\ref{Fig:Results4}, plays a significant role for the emitters grown on MgO. The interface transmission can be correlated to the lateral profile of the Fe/Pt interface. The relaxed lattice for the sample homogeneously grown at 300\,$^\circ$C possesses large undistorted Fe/Pt interface regions with lateral sizes of 20\,nm. In contrast, the stressed sample contains undistorted Fe/Pt interfaces of around 4\,nm lateral size. Such a lateral interface profile influences the ability of the interface to transfer spin angular momentum into Pt. Additionally, the induced disorder at the interface can partially depolarize and dissipate the spin current in Pt~\cite{PhysRevLett.117.207204,PhysRevLett.119.017202,PhysRevLett.112.106602} which implies further decrease of the THz emission (which is not accounted for in the context of our model). The interfacial nature of the observed dependence of the THz emission can be further confirmed in similar Fe/Pt samples with a Cu interlayer~\cite{Papaioannou2018}. The insertion of a Cu interlayer has again no influence on the spectral composition of the generated THz pulse~\cite{Papaioannou2018}. Only the amplitude is scaled down by a factor of 2 due to the reduction of the interface transmission of the two Cu interfaces. Similarly, in Fig.~\ref{fig:ExpTHz}\,\textbf{d}, the concept of interface transmission based on the degree of epitaxy is valid even for much thicker layers Fe(12\,nm)/Pt(6\,nm) of similar interface quality~\cite{Keller2018}. 

 \begin{figure*}[t]
 	\centering
	\includegraphics[width =0.75 \columnwidth]{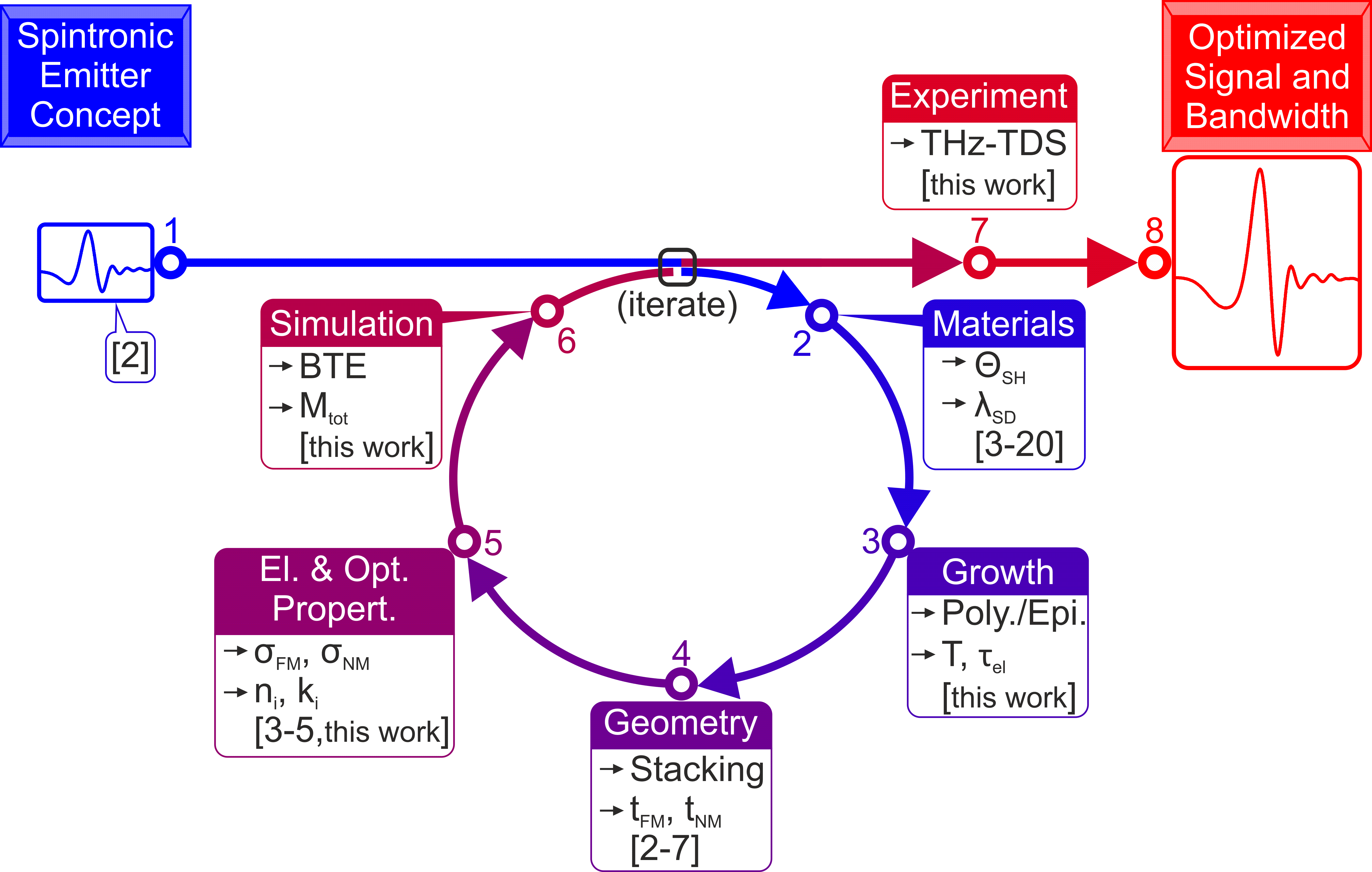}
	\caption{\label{fig:Roadmap} Roadmap to efficient spintronic THz-emitters with high signal strength and broad bandwidth. Here, $\Theta_\text{SH}$ is the spin-Hall-angle, $\lambda_\text{SD}$ is the spin-diffusion-length, $T$ and $\tau_\text{el}$ are the interface transmission parameter and the inelastic scattering lifetime, respectively. $t_\text{FM}$, $t_\text{NM}$, $\sigma_\text{FM}$ and $\sigma_\text{FM}$ are the FM and NM layer thickness and electrical conductance, respectively. $n_\text{i}$ and $\kappa_\text{i}$ are the index of refraction and the absorption coefficient of all involved layers (metalls and insulators) in the THz-range, respectively. $M_\text{tot}$ is the transfer matrix for the total layer stack.}
\end{figure*}

\subsection*{Discussion} The presented data chart the features of the emitted field amplitude and bandwidth of the THz radiation of spintronic emitters with varying growth conditions. We control the emitted THz spectra by modifying the defect density that results in changing the elastic electron-defect scattering lifetime in Fe and Pt and the interface transmission for spin-polarized, non-equilibrium electrons.
We experimentally address the aforementioned factors by studying relaxed epitaxial, deformed epitaxial and fully non-epitaxial Fe/Pt bilayers. A decreased defect density increases the electron-defect scattering lifetime and our theoretical model predicts that this results in a significant enhancement of the THz-signal amplitude and shifts the spectrum towards lower THz frequencies. The parameter of the interface transmission is correlated to the ability of the interface to transfer hot carriers into the NM layer. The transmission influences the spectral amplitude of the emitted THz field but conserves the composition of the spectrum.
Figure~\ref{fig:tau-T} provides an overview of these two effects. We plot the simulated emission properties of the spintronic emitters with respect to their scattering lifetime and to the interface transmission parameter. The size of the THz-pulse amplitude and the magnitude of the spectral width (FWHM) of the THz spectrum for different scattering lifetimes $\tau_\text{el}$ is shown in Fig.~\ref{fig:tau-T} (\textbf{a, c}). For different transmission values, $T$, we extract the same quantities, cf.~Fig.~\ref{fig:tau-T} (\textbf{e, g}). Three measurement schemes are presented: directly from the Pt side, through the MgO substrate and through the Al$_2$O$_3$ substrate. Figure ~\ref{fig:tau-T} predicts the properties of the THz emission. Furthermore, the anticipated emission spectra that account for the detector response are shown in the lower panel of Fig.~\ref{fig:tau-T} (\textbf{b, d, f, h}). The simulation results indicate that the spectral response of the detector dramatically decreases the ability to measure shifts of the spectral width in experiment.

We now turn the discussion to our experimental data. The theoretical values of Fig.~\ref{fig:tau-T} suggest that the effect of interface transmission and varying lifetimes can be disentangled once the spectral width is obtained. Again, the spectral width of the THz radiation only changes for different lifetimes, see Fig.~\ref{fig:tau-T} (\textbf{c, d}), whereas its amplitude varies for different $\tau_\text{el}$ and $T$. For deconvolution, we determine the bandwidth (FWHM) of the THz spectrum for different $\tau_\text{el}$ values and analyze the amplitude dependence on $\tau_\text{el}$ and $T$. Results for the model setup that correspond to the experiment are shown in Fig.~\ref{fig:tau-T} (\textbf{b, d, f, h}). We first discuss the case of samples grown on  different MgO and Al$_2$O$_3$ substrates. The FWHM of 1.3\,THz and 1.6\,THz for MgO and Al$_2$O$_3$ respectively from Fig.~\ref{fig:ExpTHz}\, (\textbf{a, b},inset) corresponds to differences in $\tau_\text{el}$ between the samples in the range of 30~$\mathrm{fs}$. Next, we address the amplitude ratio. A ratio A$_{non-epitaxial}$/A$_{epitaxial}$ of about 10\,\% is obtained for measurements through the substrate. 

Fig.~\ref{fig:tau-T}\,\textbf{b} suggests that such big changes in the amplitude could be mainly justified with large differences in $\tau_\text{el}$  which can be in the range of 40-50\,fs (for MgO substrate) and around 10-20\,fs for the sample on Al$_2$O$_3$ substrate. By comparing both samples in measurements from the Pt-side (see methods), however, we do not find the change in the spectral width. The absence of the Si-lens for this type of measurement can cause this effect. Even in the case that the shift is not detectable without the Si-lens, this indicates small shifts. The latter points to an additional influence of the interface transmittance, which changes the amplitude but not the spectral shape. Thus, we cannot completely disentangle both effects for this particular emitter pair as suggested by the theory. Nonetheless, both experiment and theory agree on the fact that the epitaxial structure yields higher peak-field values in all cases.

For the epitaxial-partially epitaxial Fe/Pt samples grown on MgO, no change in the spectral width (around 1.3\,THz) is observed, Fig.~\ref{fig:ExpTHz}\, (\textbf{c},inset) for all measurements with and without the Si-lens. 

The experimentally measured relative amplitude change A$_{partial-epitaxial}$/A$_{epitaxial}$, is around 50\,\% for the measurement through the substrate (see Fig.~\ref{fig:ExpTHz}\textbf{c}). If we assume the non-shift is due to  small differences in elastic scattering lifetime then the significant reduction of the amplitude should be attributed to the interface transmission. We thus assume the fully-epitaxial and partially-epitaxial sample to show the same lifetime within the layers, as the prominent difference is attributed to the nm-scale changes in the quality of the interface that has been revealed by the TEM images. By referring then to to Fig.~\ref{fig:tau-T} \textbf{h} (that is plotted for a specific $\tau_\text{el}$ = 30\,fs) the change of 60\% can be justified for samples with 100\% and 50\% transmittance.

Here, measurements from the Pt-side (see methods) underline our experimental result and further show that - in agreement with theory - the relative amplitude change is slightly reduced when not measuring through the substrate.

In conlusion, our study provides a qualitative picture of how the structural parameters, which can be related to our theoretical results, alter the spectral amplitude. The theory suggests changes in the spectral width at varying frequencies for different measurements, such that the proposed scheme to disentangle the contributions poses challenges on the experimental set-up.

In Fig.~\ref{fig:Roadmap} we sketch a roadmap of THz emission from magnetic films. The roadmap aims to predict the temporal and spatial evolution of the spin current inside the metallic layers by taking into account the generation and optical propagation of the THz wave, and to forecast the THz-pulse shapes and spectra by taking into account the electron scattering lifetime and the interfacial spin current transport.
The present experimental and theoretical study proves that among the investigated Fe/Pt-structures, a defect-free epitaxial emitter yields the highest emitted-field amplitude. Theory suggests that the spectral width can be further controlled by introducing defects in the sample, which decrease the elastic scattering time at the cost of lower peak-field amplitudes.


\section*{Materials and Methods}

\subsection*{Growth of Fe/Pt bilayers}

Fe thin films were grown epitaxially on MgO (100) substrates by electron-beam evaporation technique in an ultrahigh vacuum (UHV) chamber with a base pressure of 5 $\times$ 10$^{-11}$ mbar. The growth rate was R = 0.04 $\text{\AA}$/s controlled by a quartz crystal during the deposition procedure. The incident Fe beam was perpendicular to the MgO substrate. The cleaning protocol of the MgO (001) 1 $\times$ 1 cm$^{2}$ substrates involved annealing at 650 \,$^\circ$C, and plasma-etching processing by a 50-50 \% mix of Ar and O$_{2}$ gas. The deposition of Fe was performed at room and at 300 \,$^\circ$C substrate temperatures. After the deposition of Fe, annealing at the corresponding growth temperature was performed. At a next growth stage, a Pt layer was deposited on top of the Fe layer at 300 \,$^\circ$C or at room temperature. Layer thicknesses of Fe (2 and 12 nm)/ Pt (3 and 6 nm) were monitored in-situ by a calibrated quartz crystal oscillator and confirmed ex-situ by X-ray reflectivity (XRR) measurements.

\subsection*{Energy-filtered transmission electron microscopy}

Analysis by X-TEM was performed using a Jeol 2010 transmission electron microscope (thermionic LaB$_6$ cathode) in analytical configuration equipped with Gatan imaging filter GIF-863 Tridiem. The zero-loss EFTEM images were taken at an electron energy of 197\,keV (energy window of 10\,eV), a beam-convergence semi-angle ($\alpha$) of 0.42\,mrad and the collection semi-angle $\beta$) of 13\,mrad in the bright-field imaging mode. The TEM samples were prepared employing a common cross-section (X-TEM) procedure encompassing an ultrasonic cutting of two 2$\times$3 mm$^{2}$ plates from the sample, face-to-face gluing, mechanical grinding, and polishing, as well as a low-angle (4 \,$^\circ$) ion thinning (4 and 3\,kV) and polishing (1.5 and 1\,kV) by Ar-ions as the final steps.
The penultimate step was continued until a small (length of about some tens of $\mu$m) lens-like hole appeared at the glued interface between two sample pieces. As a result, every X-TEM sample actually contained up to four thinned sample regions (4 edges), which were further analyzed independently. The regions available for HR EFTEM extended along the sample surface up to some $\mu$m starting from each edge of the central hole. Since the thinned sample regions are separated by several tens of $\mu$m along the surface from every side of this sandwich and by several mm between the glued sample pieces, the procedure ensured that the performed EFTEM investigations were representative for the entire layers of interest. The X-TEM samples were oriented relative to the primary electron beam such that the (200) atomic planes of MgO could be directly resolved at high magnifications (HR EFTEM). An identification of the Pt and Fe individual layers was carried out by measuring the corresponding lattice spacing (HR-EFTEM) as well as by EELS elemental mapping of Mg ($K$), Fe ($L_{2,3}$) and Pt ($M_{4,5}$) edges ($\alpha=1.46$\,mrad, $\beta$=26\,mrad). For visual comparison purposes, the X-TEM sample regions of a practically equal cross-section thickness were chosen.
The Fe/Pt layers grow epitaxially on MgO\,(100) substrates following the Bain epitaxial orientation that correlates the growth of an  \textit{fcc} lattice (Pt) on top of a \textit{bcc} lattice (Fe). The Fe lattice is in-plane rotated by  45$^\circ$ with respect to both MgO and Pt providing an epitaxial growth of the three lattices.
To compare the influence of epitaxial growth onto the THz generation, we have grown the same Fe (2\,nm)/Pt (3\,nm) layers onto a sapphire, Al$_2$O$_3$\,(0001) substrate at the same conditions (uniformly at 300\,$^\circ$C). The trigonal structure of the Al$_2$O$_3$ substrate induces a large lattice mismatch with the cubic lattices of the Fe and Pt layers and leads to polycrystallinity.

\subsection*{Spintronic terahertz time domain spectroscopy} 
\begin{figure}
	\centering
	\includegraphics[width =0.6 \columnwidth]{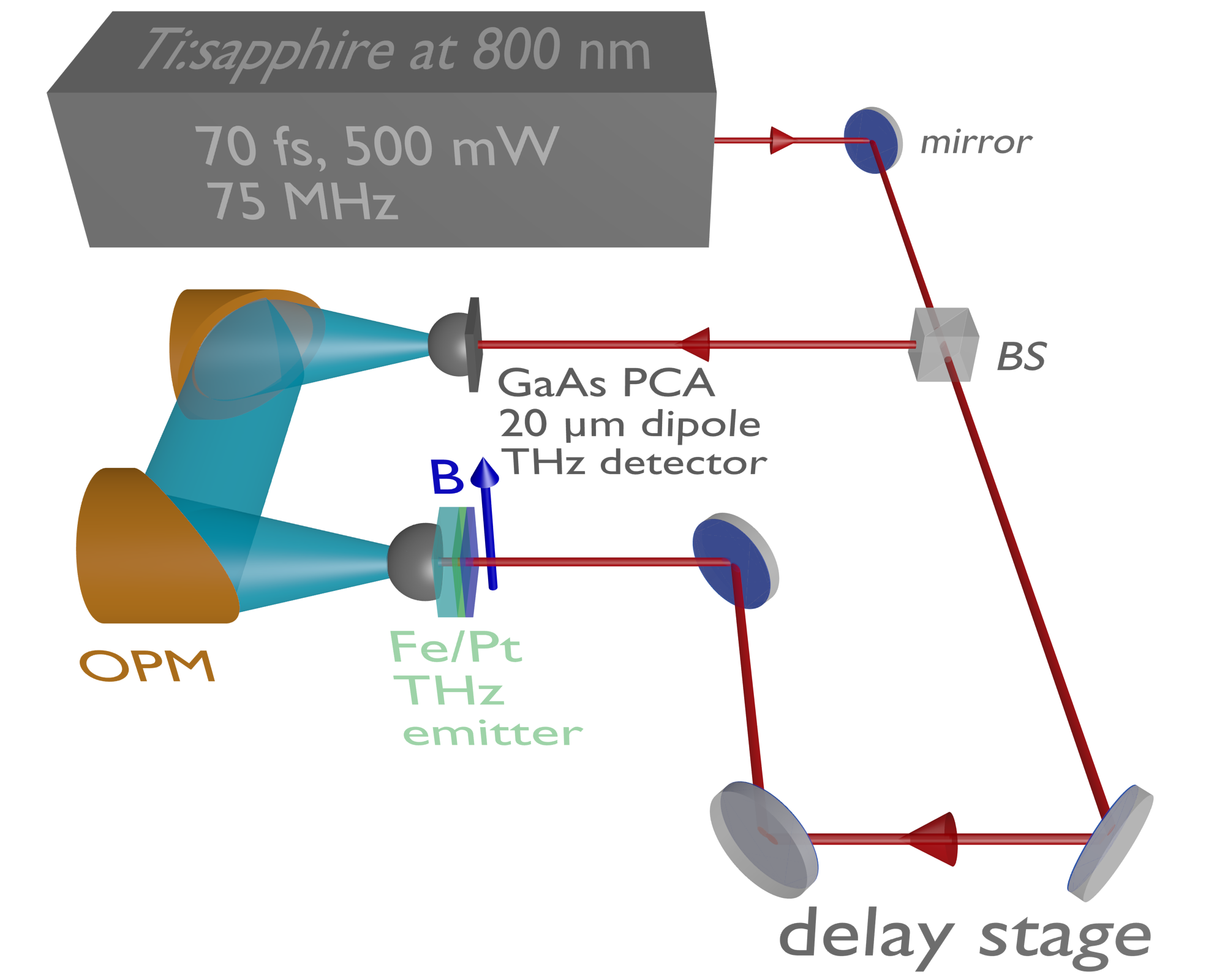}
	\caption{\label{fig:012} Illustration of the terahertz time domain spec\-tro\-sco\-pe (THz-TDS) with a spintronic heterostructure as THz-emitter.}
\end{figure}

The THz experiments with the Fe/Pt heterostructures were performed with a standard terahertz time domain spectroscopy (THz-TDS) system (Fig.~\ref{fig:012}), where the heterostructures were used as THz emitters (the performance of the system is described in detail in Ref.~\citenum{Torosyan2018}). 

The core of the system is a femtosecond Ti:Sa laser which produces 70\,fs optical pulses at a wavelength of 800\,nm with a repetition rate of 75\,MHz and a typical output power of 500\,mW. The laser beam is then split by a beam-splitter (BS) into pump and probe beam with a power ratio of 90\,:\,10. The pump beam is led through a computer-controlled delay stage onto a spintronic THz emitter and the probe beam is used to excite a photoconductive antenna with a dipole length of 20\,\textmu m that acts as THz detector. The spintronic emitter is magnetized by an external magnetic field (20\,mT) in the magnetic easy axis direction, which is perpendicular to the direction of the incident pump beam. The external field determines the polarization plane of the generated THz wave. The optical pump beam is focused onto the heterostructure by an aspherical short-focus lens. The Fe/Pt bilayer emits THz pulses into the free space as a strongly divergent beam. The pump beam is focused onto the emitter from the Pt side and a hyperhemispherical Si-lens is attached to the substrate of the emitter to collimate the beam. The so formed conical THz beam is led via two off-axis parabolic mirrors (OPM) in an f-2f-f arrangement onto a second Si-lens attached to the PCA detector. To guarantee comparable experimental conditions, the alignment of the THz optics and of the detector is not changed during the exchange of the spintronic emitters. Since the lateral layer structure of the heterostructures is homogeneous and the position of the pump beam focus stays constant, the exchange of emitters does not influence the THz-signal. With the delay stage, the arrival of the THz pulse and the probe beam pulse is synchronized and the detected voltage at the PCA can be scanned. This voltage is proportional to the momentary electric field amplitude of the THz wave and, therefore, the THz-E-field and its phase can be measured as a function of the delay time. The voltage is measured by lock-in amplification while the pump beam is optically chopped. The THz spectral amplitude can be obtained by Fourier analysis. The bandwidth of the PCA detector with the 20 \textmu m dipole length is limited to a minimum frequency of 100\,GHz and a maximum frequency of 8 THz. While the lowest measurable frequency is only limited by the dipole length of the PCA (longer dipole metalizations allow for the detection of lower frequencies), the detection of higher frequencies is limited by the strong phonon resonances of the GaAs substrate material of the PCA (absorption of the THz radiation).

\subsection*{Influence of substrate and Si lens on THz emission}

\begin{figure}
	\centering
	\includegraphics[width = 1.0 \columnwidth]{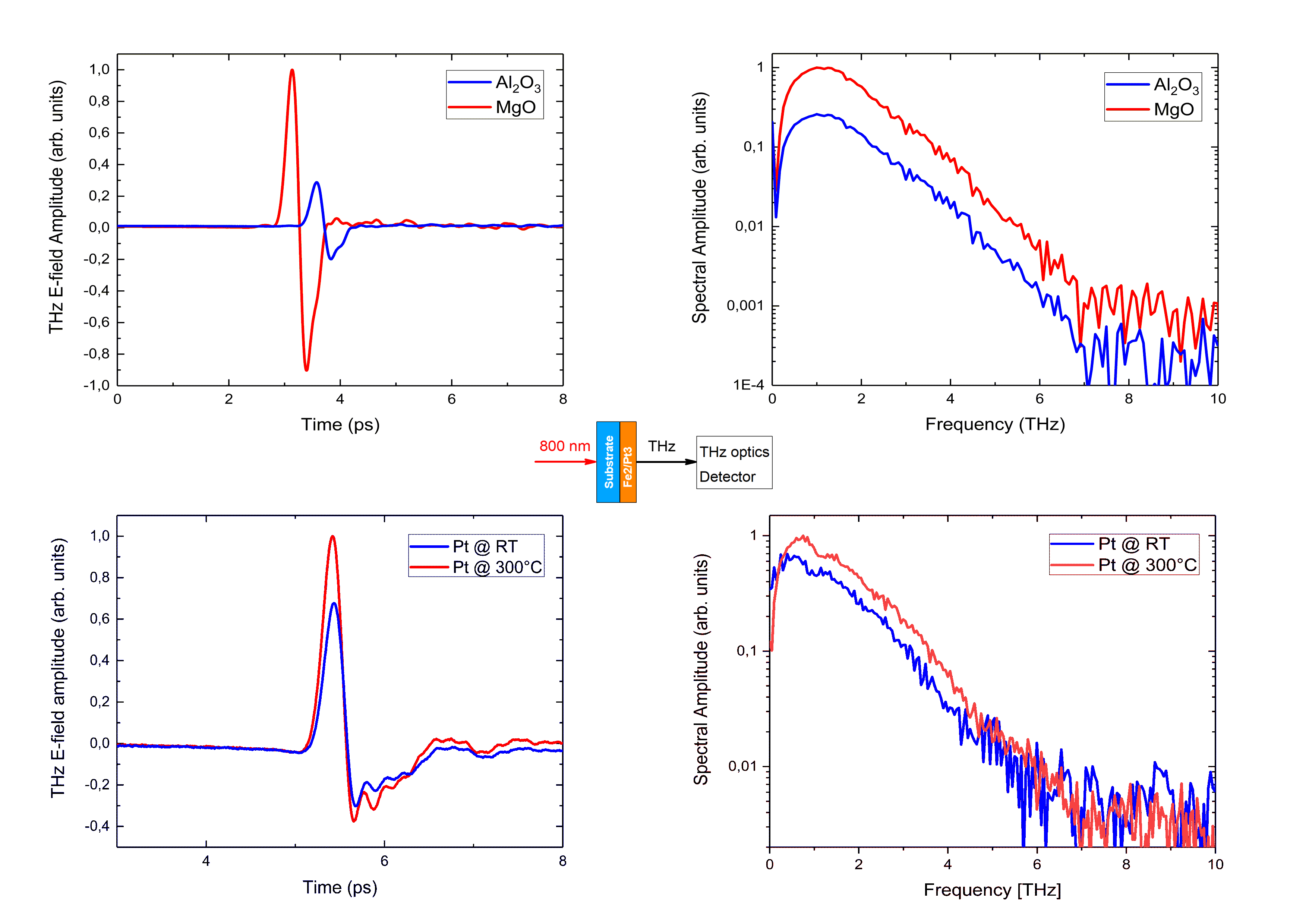}
	\caption{\label{fig:PtsideMgO} Measurements of THz emission directly from the Pt side for  Fe\,(2\,nm)\,/\,Pt\,(3\,nm) samples grown on MgO and Al$_2$O$_3$ substrates (upper panel), and for Fe\,(2\,nm)\,/\,Pt\,(3\,nm) grown on MgO, but at different temperatures as  presented in Fig.~\ref{fig:TEM0} and Fig.~\ref{fig:TEM} (lower panel). The measurements have been performed  without the presence of a Si-lens. The result undoubtedly proves that even after removing the optical influences due to the substrate and the Si-lens, the intrinsic metal/interface properties define the properties of the spintronic THz sources. The large change in the amplitude is present in all cases and for all measurements schemes, with and without the Si-lens, pumping from the substrate or from the Pt side. The temporal delay between the pulses from emitters on MgO and Al$_2$O$_3$ is due to the different refractive indices of the substrates for infrared light.}
\end{figure}

In our experimental set-up the generated THz pulse has to propagate through the substrate material and the Si lens. In order to remove the influence of substrate and Si lens we have also measured  THz emission directly from the Pt side of the bilayer without the collimating Si lens. The results shown in Fig.~\ref{fig:PtsideMgO}, demonstrate that the emission of the THz pulses is not determined by external effects (substrate, Si lens). The large change in amplitude is present in all experimental cases indicating that the metal interface properties define the characteristics of spintronic THz generation. 

\subsection*{Boltzmann transport equation}

On the femto- to picosecond timescale, two main types of scattering processes are considered in the Boltzmann transport approach: scattering events with impurities, defects and phonons are considered elastic due to the small change in energy. Secondary-carrier generation due to hot-electron scattering with equilibrium carriers is included using inelastic electron scattering times~\cite{PhysRevLett.105.027203,PhysRevLett.116.196601,Kaltenborn:2014du}. The Boltzmann transport theory has been proven to be an adequate tool to simulate excited carrier dynamics in metallic structures on the nanoscale~\cite{Hurst:2018bz,Manfredi:2005ba}. 

To arrive at a form of the BTE that is feasible to solve, we apply a number of simplifications and use the approach presented in Ref.~\citenum{Nenno2018PIC}. Since the structures are homogeneous in $x$- and $y$-direction and the length of the slab exceeds the laser-spot size, any spatial dependence in those directions can be neglected. Further, the propagation angle of the excited carriers is assumed to be uniformly distributed and all scattering processes involved are assumed to maintain this distribution. This is according to the random-$k$-approximation in Ref.~\citenum{Penn:1985wt}. There is no influence of internal fields, if small excitation densities are assumed. Additionally, any long-range fields are screened within 1\,nm~\cite{Zhu2014}. 
Thus, we model the evolution of the hot-carrier distribution function $g$ according to~\cite{Nenno2018PIC}

\begin{equation}
\begin{split}
\bigg[\partial_t + v_\sigma(E)\cos(\theta) \frac{\partial}{\partial z}\bigg] g_{\sigma}(z,E,\theta,t)~=~  S_{\sigma}(z,E,t) -\frac{g_{\sigma}(z,E,\Theta,t)}{\tau^{\mathrm{eff}}_{\sigma}(z,E)} \\  +\sum_{\sigma'}\int\frac{d\Omega'}{4\pi}\int dE' \,w(\sigma',E';\sigma,E)g_{\sigma'}(z,E',\theta',t)\rho_{\sigma'}(E') \,.
\label{eq:bte2}
\end{split}
\end{equation}

The distribution depends on spin $\sigma$, position along the axis of material stacking $z$, particle energy above the Fermi level $E$, the propagation angle $\theta$ with respect to $z$ and time $t$. The polarization axis along which the spin aligns parallel or antiparallel is given by the effective magnetic field. The left-hand side of Eq.~\ref{eq:bte2} describes the spatial and the temporal evolution of the distribution function. On the right-hand side, $S$ denotes a source term to describe the laser excitation process, which lifts carriers from below the Fermi energy to above. The latter two terms in Eq.~\ref{eq:bte2} describe out- and in-scattering processes due to interactions with equilibrium carriers and many-particle excitations. In particular, they describe the spin-dependent elastic scattering (electrons with defects\,/\,impurities) and inelastic scattering of excited electrons with bound electrons below the Fermi energy~\cite{Zhukov,Kaltenborn:2014du}. The effective, spin-dependent lifetime is given by $\tau^{\mathrm{eff}}_{\sigma}(E)=1/(\tau^{-1}_{\mathrm{el}}+\tau^{-1}_{\sigma}(z,E))$, combining the elastic lifetime $\tau_{\mathrm{el}}$ and the inelastic scattering lifetime $\tau_{\sigma}(z,E)$. The scattering amplitude $w$ controls the particle redistribution in momentum space depending on spin and energy. All scattering events are assumed to be local. The transmission $T$ of electrons at the metal/metal interface is solved stochastically in the transport term on the left side of Eq.~\ref{eq:bte2}.

\begin{figure}[t]
	\centering
	\includegraphics[width =0.55 \columnwidth]{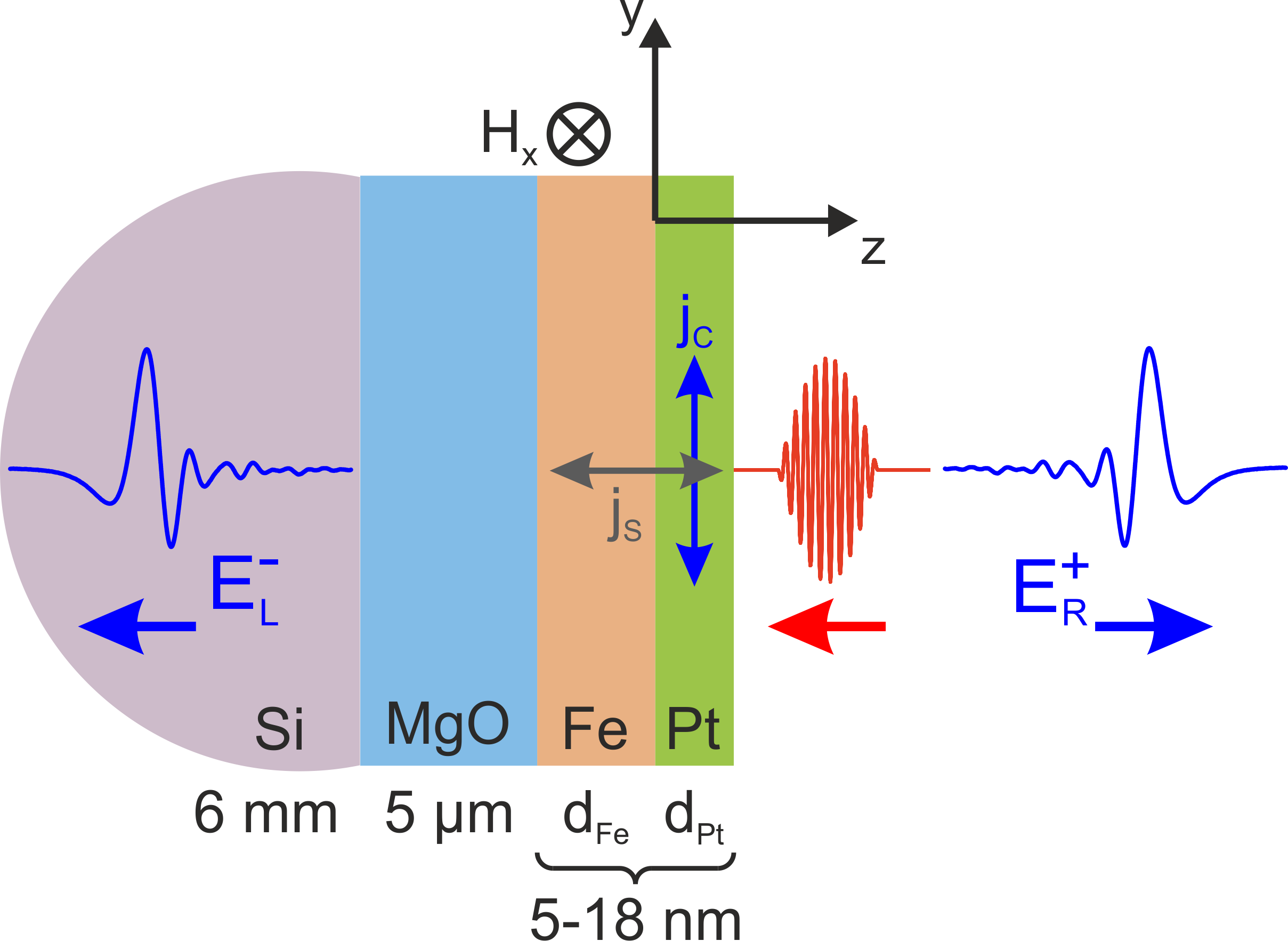}
	\caption{\label{fig:010} Schematic of the material stacking involved in the optical calculations. The fs pulse (in red) is incident from the right side. The left part of the THz wave (E-field amplitude $E_\text{L}^{-}$ according to Eq.~\ref{eq:left}) propagates the Si-lens and travels to the detector, while the right part ($E_\text{R}^{+}$) leaves the TDS setup undetected.}
\end{figure}

From the hot-electron distribution, the spin-current density $\mathbf{j}_\text{s}(z,t)$ can be calculated by

\begin{equation}
\mathbf{j_s}(z,t) = \int \frac{d\Omega}{4\pi}\int dE\,v(E)\cos(\theta) \left[g_{\uparrow}(z,E,\theta,t)-g_{\downarrow}(z,E,\theta,t)\right]\mathbf{\hat{z}}\, ,
\label{eq:spincurrent}
\end{equation}

where the integration over $\Omega$ includes all solid angles. Similar relations were first used in Ref.~\citenum{Kampfrath2013}. We use a different numerical scheme presented in Ref.~\citenum{Nenno_arxiv_THz} to calculate the induced charge current. Due to the inverse spin Hall effect in the Pt layer, this spin-current induces an effective charge current perpendicular to itself and the magnetization axis ($\sigma$-axis) by

\begin{equation}
\mathbf{j}_\text{c} = \Theta_{\text{SH}}~\mathbf{j}_\text{s} \times \frac{\mathbf{M}}{|\mathbf{M}|}\,,
\label{eq:spinhall}
\end{equation}

where the spin Hall angle is denoted by $\Theta_{\text{SH}}$ and it is assumed that the magnetization points in positive $x$-direction, simplifying Eq.~\ref{eq:spinhall} to $j_\text{c} = \Theta_{\text{SH}}~{j}_\text{s}$, with ${j}_\text{c}$ as magnitude in $y$-direction and ${j}_\text{s}$ in $z$-direction, along the layer stacking (see Fig.~\ref{fig:010}). For that, we assume a spin-Hall angle of 0.06~\cite{Keller2018} for the conversion efficiency from spin- to charge current.
We solve Eq.~\eqref{eq:bte2} using a Particle-In-Cell technique combined with the Strang operator splitting~\cite{Faghihi:2017vv,Nenno2018PIC}. Carrier velocities at different energies as well as their lifetimes are taken from Ref.~\citenum{Kaltenborn:2014du} for Fe and from Ref.~\citenum{Zhukov} for Pt. There are no fit-parameters in this model besides the elastic scattering time and the transmission function, as described in the main text, as all material data is extracted from \textit{ab initio} calculations.

\begin{figure}[t]
	\centering
	\includegraphics[width =0.5 \columnwidth]{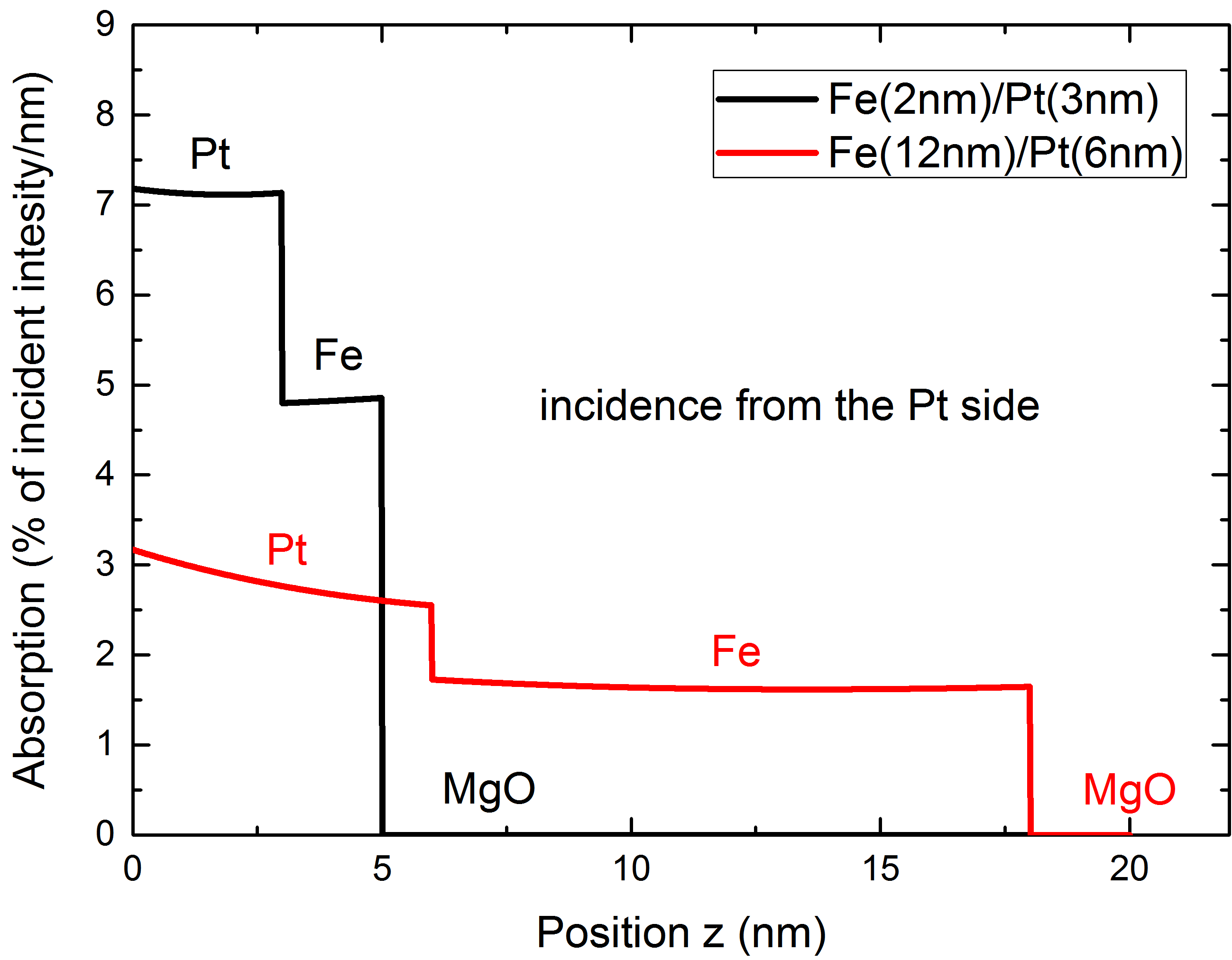}
	\caption{Absorption of the fs pulse through the Fe(12\,nm)/Pt(6\,nm) and Fe\,(2\,nm)\,/\,Pt\,(3\,nm) bilayers with incidence from the Pt side. \label{Fig:Laser_Absorption}}
\end{figure}

\subsection*{Spatial profile of the laser absorption}
\label{sec:laser_absorption}

The absorption of the laser pulse in the considered structures is calculated using a numerical model to solve the one-dimensional wave equation for s-polarized light. The absorption profile enters the numerical calculation by comparing the number of excited particles to the absorbed laser energy, depending on position~\cite{PhysRevLett.85.844}. Optical parameters at 800\,nm wavelength are taken from Ref.~\citenum{Werner:2009gt} for Fe and Pt and from Ref.~\citenum{stephens1952index} for MgO. The fraction of absorbed incident laser power over position is plotted in Fig.~\ref{Fig:Laser_Absorption}. The total absorption is 36\,\% and 41\,\% for a MgO(0.5mm)/Fe(12\,nm)/Pt(6\,nm) and MgO(0.5\,mm)/Fe\,(2\,nm)\,/\,Pt\,(3\,nm) heterostructure, respectively, where the laser pulse is incident on the Pt surface. Similar results are found for the fs laser absorption of the emitter grown on sapphire (not shown).

\begin{figure*}
	\centering
	\includegraphics[width =1.0 \columnwidth]{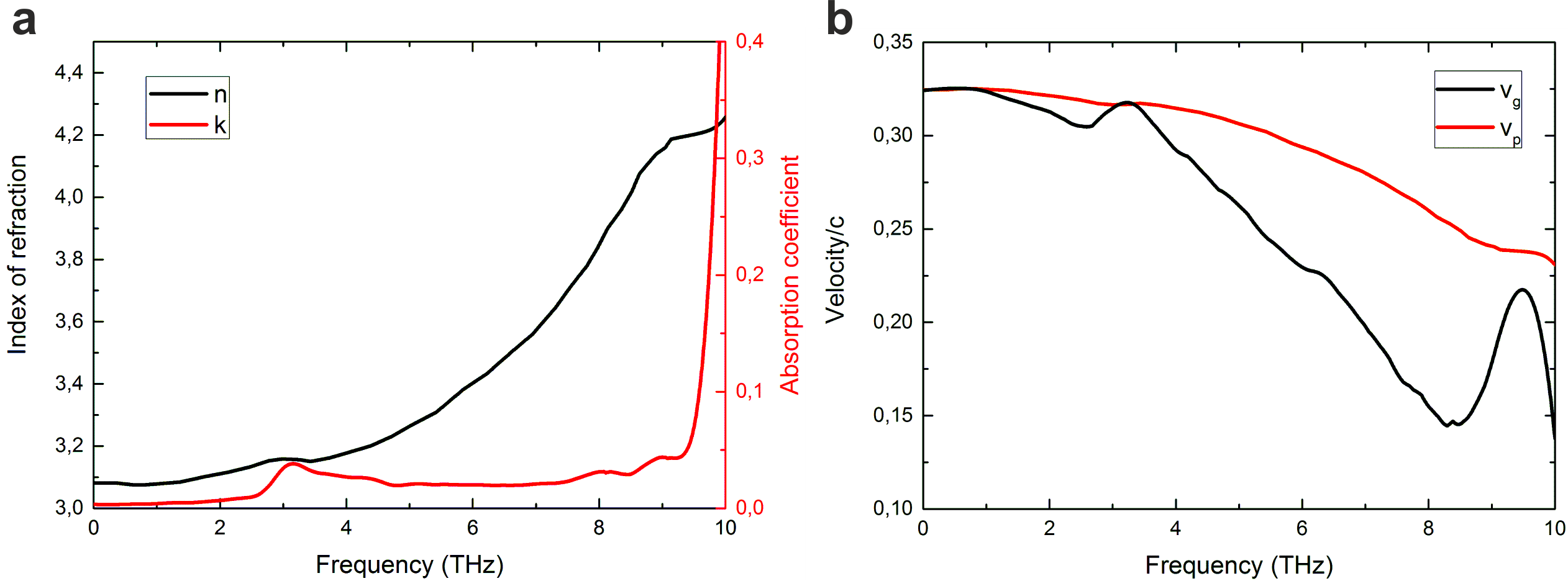}
	\caption{\textbf{a} Refractive and absorption index of MgO in the THz range. \textbf{b} Group and phase velocity in MgO depending on frequency using a Lorentz Oscillator model for the dielectric function~\cite{RenGuanhua:825001}. \label{Fig:Group_Phase_Velocity_MgO}}
\end{figure*}

\subsection*{Optical parameters of the substrate in the terahertz range}
\label{sec:optics_substrate}

For a realistic description of the THz emission and propagation within the structure,  we incorporate the material-dependent dielectric functions for the three main layers. The refractive index $n$ and the absorption coefficient $\kappa$ in MgO are calculated using fit data from experiments for a Lorentz oscillator model as presented in \cite{RenGuanhua:825001}. The results for MgO are shown in Fig.~\ref{Fig:Group_Phase_Velocity_MgO}\,\textbf{a}. For the Fe and Pt layer, a Drude fit yields satisfactory results in the spectroscopic range considered~\cite{Ordal:85}. In addition, it is found that changing the optical parameters in the two metals does not change the results substantially due to their small optical thickness. It is not within the scope of this work to model the silicon lens mounted on top of the MgO layer and propagation effects therein. However, the transition from MgO to Silicon is included assuming a constant index of refraction in silicon of $n_{\mathrm{Si}}\approx$\,3.42~\cite{Chandler-Horowitz}. For the 0.5\,mm-thick sapphire (Al$_2$O$_3$) substrate, a constant refractive and absorption index of 3.33 and 0.01 are assumed, respectively, and constant phase/group velocity of 0.30/$c_0$ (ordinary beam) up to a frequency of 6\,THz for the here-discussed spectra~\cite{Querry1985}.

Since the THz waves must propagate through a relatively thick layer of MgO, it is instructive to investigate the group- and phase velocity in this material. The resulting curves using the Lorentz Oscillator model~\cite{RenGuanhua:825001} are shown in Fig.~\ref{Fig:Group_Phase_Velocity_MgO}\,\textbf{b}. Both curves show a negative slope, showing that higher frequencies need longer times to propagate through the slab (dispersion).

\subsection{Wave equation and transfer-matrix method}
\label{sec:helmholtz}

To calculate the THz optical field measured in the experiment, we use the inhomogeneous wave/Helmholtz equation,~\cite{Binder_book}

\begin{equation}
\left\{ \frac{\partial ^{2}}{\partial z^{2}}+\frac{\omega ^{2}}{c^{2}}%
(\varepsilon ^{\prime }+i\varepsilon ^{\prime \prime })\right\} E(z,\omega
)=- \frac{\omega ^{2}}{\epsilon_0 c^{2}}P(z,\omega )\, ,
\label{eq:helmholtz}
\end{equation}

where $\omega$ denotes the photon frequency, $\epsilon_0$ the vacuum permittivity, $c$ the speed of light in vacuum and $z$ denotes the direction normal to the layers in the heterostructure. The optical properties of the materials involved are given by the real and imaginary part of the dielectric function, $\varepsilon'$ and $\varepsilon''$, respectively, which are dependent on the photon frequency as well. The relation between those quantities and the wave vector for each material is given by:

\begin{equation}
k=\frac{\omega }{c}\sqrt{\varepsilon ^{\prime }+i\varepsilon ^{\prime \prime
}}=\frac{\omega }{c}(n+i\kappa )=\frac{\omega }{c}\widetilde{n} \, ,
\label{eq:k}
\end{equation}%

where $\widetilde{n} = n+i\kappa $ is the complex index of refraction with its real part $n$ and the absorption coefficient $\kappa$. The laser-induced charge current is related to the electric polarization $P$ via

\begin{equation}
P(z,\omega)=i\frac{j_\text{c}(z,\omega)}{\omega} \, .
\label{eq:Ej}
\end{equation}

It serves as a source for terahertz radiation. The Green's function to this problem is defined by

\begin{equation}
\left\{ \frac{\partial ^{2}}{\partial z^{2}}+k^{2}\right\} G(z,z^{^{\prime
}})=\delta (z-z^{\prime }) \, ,
\end{equation}

and can be analytically solved by the following Green's function

\begin{equation}
G(z-z^{\prime })=\frac{-i}{2k}e^{\pm ik|z-z^{\prime }|} \, .
\end{equation}%

The particular solution can then be constructed via

\begin{equation}
E_{\mathrm{inh}}(z)=i\frac{\omega  }{2\epsilon_0 c\widetilde{n}}\int dz^{\prime
}e^{\pm ik\left\vert z-z^{\prime }\right\vert }P(z^{\prime }) \, .
\label{eq:EP}
\end{equation}%

In good approximation, the spin current is found to be constant throughout the emitting Pt layer. In addition, all metallic layers are very thin, such that $P(z,\omega)=P_{0}(\omega)$ can be assumed independent of $z$ in these layers.

The transfer matrix formalism is used to describe the propagation and transmission/reflection of the generated THz wave throughout the layers. The transfer matrix for the propagation inside each layer is given by:
	
\begin{equation}
M^{(d)}=\left( 
\begin{array}{cc}
e^{ikd} & 0 \\ 
0 & e^{-ikd}%
\end{array}%
\right)\, ,
\end{equation}%

and the step matrix for each interface from left (L) to right (R) is

\begin{equation}
M_{n_{L},n_{R}}^{\mathrm{step}}=\frac{1}{2n_{R}}\left( 
\begin{array}{cc}
n_{R}+n_{L} & n_{R}-n_{L} \\ 
n_{R}-n_{L} & n_{R}+n_{L}%
\end{array}%
\right) \, .
\end{equation}%

The propagation of forward ($+$) and backward ($-$) traveling waves $E_\text{R}$ and $E_\text{L}$ (electric field amplitudes on the right and left side of the Pt layer) in $z$-direction is then calculated as

\begin{equation}
\left( 
\begin{array}{c}
E_\text{R}^{+} \\ 
E_\text{R}^{-}
\end{array}%
\right) =M^{\mathrm{tot}}\left( 
\begin{array}{c}
E_\text{L}^{+} \\ 
E_\text{L}^{-}%
\end{array}%
\right) +\left( 
\begin{array}{c}
E_{0}^+ \\ 
E_{0}^-%
\end{array}%
\right)\, ,
\end{equation}%

where, for the experimentally used structure, the total propagation matrix is given by

\begin{equation}
\begin{split}
M^{\mathrm{tot}}=M_{\mathrm{Pt,Vac.}}^{\mathrm{step}}M_{\mathrm{\mathrm{Pt}}}^{(d_{\mathrm{\mathrm{Pt}}})}M_{\mathrm{\mathrm{Fe,Pt}}}^{\mathrm{step}}M_{\mathrm{Fe}}^{(d_{\mathrm{\mathrm{Fe}}})}
 M_{\mathrm{MgO,Fe}}^{\mathrm{step}}M_{\mathrm{MgO}}^{(d_{\mathrm{\mathrm{MgO}}})}M_{\mathrm{Si,MgO}}^{\mathrm{step}}\, ,
\end{split}
\label{eq:M}
\end{equation}%

and the contribution of the source is found to be

\begin{equation}
\left( 
\begin{array}{c}
E_{0}^+ \\ 
E_{0}^-%
\end{array}%
\right) =  M_{\mathrm{Pt,Vac.}}^{\mathrm{step}}\left( 
\begin{array}{c}
e^{ik_{\mathrm{\mathrm{Pt}}}d_{\mathrm{\mathrm{Pt}}}}-1 \\ 
e^{-ik_{\mathrm{\mathrm{Pt}}}d_{\mathrm{\mathrm{Pt}}}}-1%
\end{array}%
\right) \frac{1 }{2\epsilon_0 \widetilde{n}_\text{Pt}^2} P_0~.
\end{equation}%

Finally, the emitted field amplitudes are obtained as

\begin{equation}
E_\text{L}^{-}=-\frac{E_{0}^-}{M_{22}^{\mathrm{tot}}}\, ,
\label{eq:left}
\end{equation}

for the backward propagating wave and

\begin{equation}
E_\text{R}^{+}=-\frac{M_{12}^{\mathrm{tot}}}{M_{22}^{\mathrm{tot}}}E_{0}^- + E_{0}^+ \, ,
\end{equation}

for the forward propagating one, where $E_\text{R}^{+}$ and $E_\text{L}^{-}$ are the electric field amplitudes for emission at the MgO/Si and Pt/Air interface, respectively. Only $E_\text{L}^{-}$ is measured by the THz detector. The subscripts of the total transmission matrix $M^\mathrm{tot}$ denote its elements. It is worth mentioning that these results have been compared to the findings of Sipe, presented in Ref.~\citenum{Sipe:87}, and a similar formulation limited to a single metallic layer was used in Ref.~\citenum{Kampfrath2013}.

\begin{figure}[t]
	\centering
	\includegraphics[width =0.55 \columnwidth]{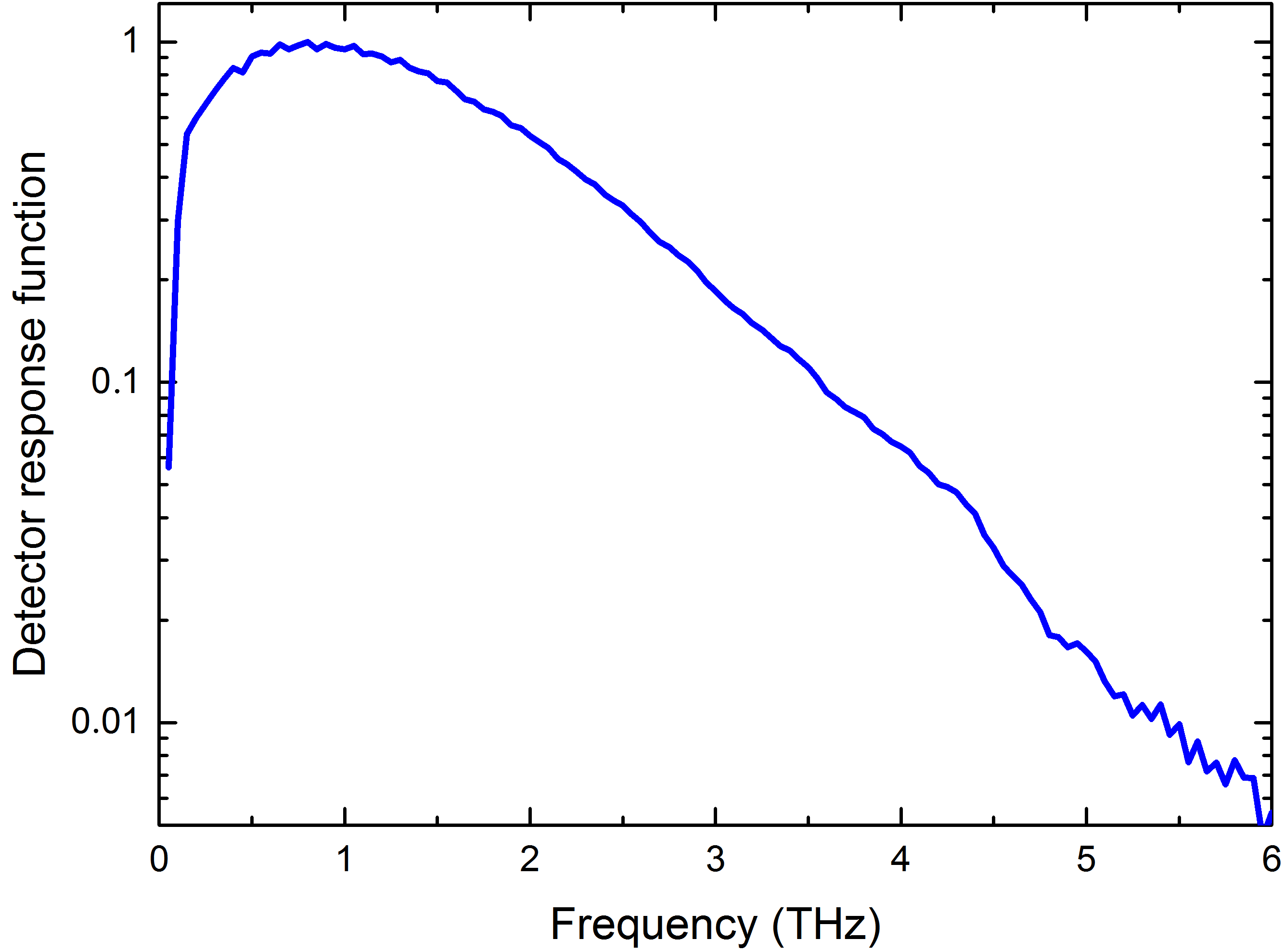}
	\caption{\label{fig:resp} Calculated detector response function of the photoconducting antenna with a 20 \textmu m dipole length used in this publication. It is obtained by comparing the simulated and experimental spectra for the Fe\,(2\,nm)\,/\,Pt\,(3\,nm) emitter grown on sapphire ($E^{\text{Al}_2\text{O}_3}_\text{exp}(\omega)/E^{\text{Al}_2\text{O}_3}_\text{sim}(\omega,\tau_\text{el}=10\,\text{fs})$ with the maximum response normalized to 1.}
\end{figure}

\subsection*{PCA detector response function}
\label{sec:response}

The BTE simulation, the Green function formalism and the transfer matrix methods together show which THz pulse shapes and spectral components arrive at the detector. Additionally to the absorption losses of the THz wave by the metallic layers and the substrate, the experimentally measured spectra are also modified by the frequency-dependent sensitivity of the photoconducting antenna used as detector. We extract the detector function by comparing the numerically simulated THz-spectra of the Fe(2nm)/Pt(3nm) emitter grown on sapphire with the experimentally obtained spectra. As shown in Fig.~\ref{fig:resp}, the detector has its largest sensitivity at a frequency of about 850\,GHz. The response of the detector has a 30\,\%-value at 100\,GHz and a steep drop at lower frequencies. For larger frequencies, the 10\,\% and 1\,\%-values are 3.5 and 5.5\,THz, respectively. The results are similar to the calculations by Jepsen et al.~\cite{Jepsen:96}. Due to the strong phonon resonances of the GaAs substrate material, the upper limit of the measurable bandwidth is around 8\,THz. Assuming the case that the THz-wave and the PCA response function have their spectral maximum at different frequencies, the maximum of detected spectra will be shifted but differences in the spectral maxima for different emitters are still visible.


\begin{thebibliography}{10}
\expandafter\ifx\csname url\endcsname\relax
  \def\url#1{\texttt{#1}}\fi
\expandafter\ifx\csname urlprefix\endcsname\relax\def\urlprefix{URL }\fi
\expandafter\ifx\csname doiprefix\endcsname\relax\def\doiprefix{DOI }\fi
\providecommand{\bibinfo}[2]{#2}
\providecommand{\eprint}[2][]{\url{#2}}

\bibitem{Walowski:2016bt}
\bibinfo{author}{Walowski, J.} \& \bibinfo{author}{M{\"u}nzenberg, M.}
\newblock \bibinfo{title}{{Perspective: Ultrafast magnetism and THz
  spintronics}}.
\newblock \emph{\bibinfo{journal}{Journal of Applied Physics}}
  \textbf{\bibinfo{volume}{120}}, \bibinfo{pages}{140901--17}
  (\bibinfo{year}{2016}).

\bibitem{Kampfrath2013}
\bibinfo{author}{Kampfrath, T.} \emph{et~al.}
\newblock \bibinfo{title}{Terahertz spin current pulses controlled by magnetic
  heterostructures}.
\newblock \emph{\bibinfo{journal}{Nature Nanotechnology}}
  \textbf{\bibinfo{volume}{8}}, \bibinfo{pages}{256} (\bibinfo{year}{2013}).
\newblock \urlprefix\url{https://doi.org/10.1038/nnano.2013.43}.

\bibitem{Torosyan2018}
\bibinfo{author}{{G. Torosyan, S. Keller, L. Scheuer, R. Beigang, and E. Th.
  Papaioannou}}.
\newblock \bibinfo{title}{Optimized spintronic terahertz emitters based on
  epitaxial grown fe/pt layer structures}.
\newblock \emph{\bibinfo{journal}{Sci. Rep.}} \textbf{\bibinfo{volume}{8}},
  \bibinfo{pages}{1311} (\bibinfo{year}{2018}).

\bibitem{Seifert2016}
\bibinfo{author}{Seifert, T.} \emph{et~al.}
\newblock \bibinfo{title}{Efficient metallic spintronic emitters of
  ultrabroadband terahertz radiation}.
\newblock \emph{\bibinfo{journal}{Nat. Photon.}} \textbf{\bibinfo{volume}{10}},
  \bibinfo{pages}{483--488} (\bibinfo{year}{2016}).

\bibitem{ADMA:ADMA201603031}
\bibinfo{author}{Wu, Y.} \emph{et~al.}
\newblock \bibinfo{title}{High-performance thz emitters based on
  ferromagnetic/nonmagnetic heterostructures}.
\newblock \emph{\bibinfo{journal}{Advanced Materials}}
  \textbf{\bibinfo{volume}{29}}, \bibinfo{pages}{1603031}
  (\bibinfo{year}{2017}).

\bibitem{ADOM:ADOM201600270}
\bibinfo{author}{Yang, D.} \emph{et~al.}
\newblock \bibinfo{title}{Powerful and tunable thz emitters based on the fe/pt
  magnetic heterostructure}.
\newblock \emph{\bibinfo{journal}{Advanced Optical Materials}}
  \textbf{\bibinfo{volume}{4}}, \bibinfo{pages}{1944--1949}
  (\bibinfo{year}{2016}).

\bibitem{Papaioannou2018}
\bibinfo{author}{Papaioannou, E.~T.} \emph{et~al.}
\newblock \bibinfo{title}{Efficient terahertz generation using fe/pt spintronic
  emitters pumped at different wavelengths}.
\newblock \emph{\bibinfo{journal}{IEEE Transactions on Magnetics}}
  \textbf{\bibinfo{volume}{54}}, \bibinfo{pages}{1--5} (\bibinfo{year}{2018}).
\newblock \doiprefix 10.1109/TMAG.2018.2847031.

\bibitem{Qiu:18}
\bibinfo{author}{Qiu, H.~S.} \emph{et~al.}
\newblock \bibinfo{title}{Layer thickness dependence of the terahertz emission
  based on spin current in ferromagnetic heterostructures}.
\newblock \emph{\bibinfo{journal}{Opt. Express}} \textbf{\bibinfo{volume}{26}},
  \bibinfo{pages}{15247--15254} (\bibinfo{year}{2018}).
\newblock
  \urlprefix\url{http://www.opticsexpress.org/abstract.cfm?URI=oe-26-12-15247}.
\newblock \doiprefix 10.1364/OE.26.015247.

\bibitem{spin2017}
\bibinfo{author}{Seifert, T.} \emph{et~al.}
\newblock \bibinfo{title}{Terahertz spin currents and inverse spin hall effect
  in thin-film heterostructures containing complex magnetic compounds}.
\newblock \emph{\bibinfo{journal}{SPIN}} \textbf{\bibinfo{volume}{07}},
  \bibinfo{pages}{1740010} (\bibinfo{year}{2017}).
\newblock \urlprefix\url{https://doi.org/10.1142/S2010324717400100}.
\newblock \doiprefix 10.1142/S2010324717400100.
\newblock \eprint{https://doi.org/10.1142/S2010324717400100}.

\bibitem{Albrecht2018}
\bibinfo{author}{Schneider, R.} \emph{et~al.}
\newblock \bibinfo{title}{Magnetic-field-dependent thz emission of spintronic
  tbfe/pt layers}.
\newblock \emph{\bibinfo{journal}{ACS Photonics}} \textbf{\bibinfo{volume}{5}},
  \bibinfo{pages}{3936--3942} (\bibinfo{year}{2018}).
\newblock \urlprefix\url{https://doi.org/10.1021/acsphotonics.8b00839}.

\bibitem{Haifeng2018}
\bibinfo{author}{Feng, Z.} \emph{et~al.}
\newblock \bibinfo{title}{Highly efficient spintronic terahertz emitter enabled
  by metal–dielectric photonic crystal}.
\newblock \emph{\bibinfo{journal}{Advanced Optical Materials}}
  \textbf{\bibinfo{volume}{0}}, \bibinfo{pages}{1800965}
  (\bibinfo{year}{2018}).
\newblock
  \urlprefix\url{https://onlinelibrary.wiley.com/doi/abs/10.1002/adom.201800965}.
\newblock \doiprefix 10.1002/adom.201800965.
\newblock
  \eprint{https://onlinelibrary.wiley.com/doi/pdf/10.1002/adom.201800965}.

\bibitem{Li_2018}
\bibinfo{author}{Li, G.} \emph{et~al.}
\newblock \bibinfo{title}{Laser induced {THz} emission from femtosecond
  photocurrents in {Co}/{ZnO}/{Pt} and {Co}/{Cu}/{Pt} multilayers}.
\newblock \emph{\bibinfo{journal}{Journal of Physics D: Applied Physics}}
  \textbf{\bibinfo{volume}{51}}, \bibinfo{pages}{134001}
  (\bibinfo{year}{2018}).
\newblock \urlprefix\url{https://doi.org/10.1088%2F1361-6463%2Faaab8f}.
\newblock \doiprefix 10.1088/1361-6463/aaab8f.

\bibitem{Seifert_2018}
\bibinfo{author}{Seifert, T.~S.} \emph{et~al.}
\newblock \bibinfo{title}{Terahertz spectroscopy for all-optical spintronic
  characterization of the spin-hall-effect metals pt, w and cu80ir20}.
\newblock \emph{\bibinfo{journal}{Journal of Physics D: Applied Physics}}
  \textbf{\bibinfo{volume}{51}}, \bibinfo{pages}{364003}
  (\bibinfo{year}{2018}).
\newblock \urlprefix\url{https://doi.org/10.1088%2F1361-6463%2Faad536}.
\newblock \doiprefix 10.1088/1361-6463/aad536.

\bibitem{Li_2019}
\bibinfo{author}{{Li}, G.} \emph{et~al.}
\newblock \bibinfo{title}{{THz emission from Co/Pt bilayers with varied
  roughness, crystal structure, and interface intermixing}}.
\newblock \emph{\bibinfo{journal}{arXiv e-prints}}
  \bibinfo{pages}{arXiv:1903.04423} (\bibinfo{year}{2019}).
\newblock \eprint{1903.04423}.

\bibitem{PhysRevLett.121.086801}
\bibinfo{author}{Zhou, C.} \emph{et~al.}
\newblock \bibinfo{title}{Broadband terahertz generation via the interface
  inverse rashba-edelstein effect}.
\newblock \emph{\bibinfo{journal}{Phys. Rev. Lett.}}
  \textbf{\bibinfo{volume}{121}}, \bibinfo{pages}{086801}
  (\bibinfo{year}{2018}).
\newblock
  \urlprefix\url{https://link.aps.org/doi/10.1103/PhysRevLett.121.086801}.
\newblock \doiprefix 10.1103/PhysRevLett.121.086801.

\bibitem{PhysRevLett.120.207207}
\bibinfo{author}{Jungfleisch, M.~B.} \emph{et~al.}
\newblock \bibinfo{title}{Control of terahertz emission by ultrafast
  spin-charge current conversion at rashba interfaces}.
\newblock \emph{\bibinfo{journal}{Phys. Rev. Lett.}}
  \textbf{\bibinfo{volume}{120}}, \bibinfo{pages}{207207}
  (\bibinfo{year}{2018}).
\newblock
  \urlprefix\url{https://link.aps.org/doi/10.1103/PhysRevLett.120.207207}.
\newblock \doiprefix 10.1103/PhysRevLett.120.207207.

\bibitem{Herapath2019}
\bibinfo{author}{Herapath, R.~I.} \emph{et~al.}
\newblock \bibinfo{title}{Impact of pump wavelength on terahertz emission of a
  cavity-enhanced spintronic trilayer}.
\newblock \emph{\bibinfo{journal}{Applied Physics Letters}}
  \textbf{\bibinfo{volume}{114}}, \bibinfo{pages}{041107}
  (\bibinfo{year}{2019}).
\newblock \urlprefix\url{https://doi.org/10.1063/1.5048297}.
\newblock \doiprefix 10.1063/1.5048297.
\newblock \eprint{https://doi.org/10.1063/1.5048297}.

\bibitem{Ilya2017}
\bibinfo{author}{Razdolski, I.} \emph{et~al.}
\newblock \bibinfo{title}{Nanoscale interface confinement of ultrafast spin
  transfer torque driving non-uniform spin dynamics}.
\newblock \emph{\bibinfo{journal}{Nature Communications}}
  \textbf{\bibinfo{volume}{8}}, \bibinfo{pages}{15007} (\bibinfo{year}{2017}).
\newblock \urlprefix\url{https://doi.org/10.1038/ncomms15007}.

\bibitem{doi:10.1021/acs.nanolett.7b04538}
\bibinfo{author}{Cramer, J.} \emph{et~al.}
\newblock \bibinfo{title}{Complex terahertz and direct current inverse spin
  hall effect in yig/cu1–xirx bilayers across a wide concentration range}.
\newblock \emph{\bibinfo{journal}{Nano Letters}} \textbf{\bibinfo{volume}{18}},
  \bibinfo{pages}{1064--1069} (\bibinfo{year}{2018}).
\newblock \urlprefix\url{https://doi.org/10.1021/acs.nanolett.7b04538}.
\newblock \doiprefix 10.1021/acs.nanolett.7b04538.
\newblock \eprint{https://doi.org/10.1021/acs.nanolett.7b04538}.

\bibitem{yigPt2018}
\bibinfo{author}{Seifert, T.~S.} \emph{et~al.}
\newblock \bibinfo{title}{Femtosecond formation dynamics of the spin seebeck
  effect revealed by terahertz spectroscopy}.
\newblock \emph{\bibinfo{journal}{Nature Communications}}
  \textbf{\bibinfo{volume}{9}}, \bibinfo{pages}{2899} (\bibinfo{year}{2018}).
\newblock \urlprefix\url{https://doi.org/10.1038/s41467-018-05135-2}.
\newblock \doiprefix 10.1038/s41467-018-05135-2.

\bibitem{Sasaki}
\bibinfo{author}{{Y. Sasaki, K.Z. Suzuki, and S. Mizukami}}.
\newblock \bibinfo{title}{Annealing effect on laser pulse-induced thz wave
  emission in ta/cofeb/mgo films}.
\newblock \emph{\bibinfo{journal}{Appl. Phys. Lett.}}
  \textbf{\bibinfo{volume}{111}}, \bibinfo{pages}{102401}
  (\bibinfo{year}{2017}).

\bibitem{Evangelos}
\bibinfo{author}{Papaioannou, E.~T.} \emph{et~al.}
\newblock \bibinfo{title}{{Optimizing the spin-pumping induced inverse spin
  Hall voltage by crystal growth in Fe/Pt bilayers}}.
\newblock \emph{\bibinfo{journal}{Appl. Phys. Lett.}}
  \textbf{\bibinfo{volume}{103}}, \bibinfo{pages}{162401}
  (\bibinfo{year}{2013}).

\bibitem{Andres}
\bibinfo{author}{{A. Conca, S. Keller, L. Mihalceanu, T. Kehagias, G. P.
  Dimitrakopulos, B. Hillebrands and E. Th. Papaioannou}}.
\newblock \bibinfo{title}{Study of fully epitaxial fe/pt bilayers for spin
  pumping by fmr spectroscopy}.
\newblock \emph{\bibinfo{journal}{Phys.~Rev.~B}} \textbf{\bibinfo{volume}{93}},
  \bibinfo{pages}{134405} (\bibinfo{year}{2016}).

\bibitem{Keller2018}
\bibinfo{author}{{S. Keller, L. Mihalceanu, M. R. Schweizer, P. Lang, B. Heinz,
  M. Geilen, T. Br{\"a}cher, P. Pirro, T. Meyer, A. Conca, D. Karfaridis, G.
  Vourlias, T. Kehagias, B. Hillebrands, and E. Th. Papaioannou}}.
\newblock \bibinfo{title}{Determination of the spin hall angle in
  single-crystalline pt films from spin pumping experiments}.
\newblock \emph{\bibinfo{journal}{New J. Phys.}} \textbf{\bibinfo{volume}{20}},
  \bibinfo{pages}{053002} (\bibinfo{year}{2018}).

\bibitem{Hurst:2018bz}
\bibinfo{author}{Hurst, J.}, \bibinfo{author}{Hervieux, P.-A.} \&
  \bibinfo{author}{Manfredi, G.}
\newblock \bibinfo{title}{{Spin current generation by ultrafast laser pulses in
  ferromagnetic nickel films}}.
\newblock \emph{\bibinfo{journal}{Physical Review B}}
  \textbf{\bibinfo{volume}{97}}, \bibinfo{pages}{014424--5}
  (\bibinfo{year}{2018}).

\bibitem{Manfredi:2005ba}
\bibinfo{author}{Manfredi, G.} \& \bibinfo{author}{Hervieux, P.~A.}
\newblock \bibinfo{title}{{Finite-size and nonlinear effects on the ultrafast
  electron transport in thin metal films}}.
\newblock \emph{\bibinfo{journal}{Physical Review B}}
  \textbf{\bibinfo{volume}{72}}, \bibinfo{pages}{155421--13}
  (\bibinfo{year}{2005}).

\bibitem{Nenno2018PIC}
\bibinfo{author}{Nenno, D.~M.}, \bibinfo{author}{Rethfeld, B.} \&
  \bibinfo{author}{Schneider, H.~C.}
\newblock \bibinfo{title}{Particle-in-cell simulation of ultrafast hot-carrier
  transport in fe/au heterostructures}.
\newblock \emph{\bibinfo{journal}{Phys. Rev. B}} \textbf{\bibinfo{volume}{98}},
  \bibinfo{pages}{224416} (\bibinfo{year}{2018}).
\newblock \urlprefix\url{https://link.aps.org/doi/10.1103/PhysRevB.98.224416}.
\newblock \doiprefix 10.1103/PhysRevB.98.224416.

\bibitem{Nenno_arxiv_THz}
\bibinfo{author}{Nenno, D.~M.}, \bibinfo{author}{Binder, R.} \&
  \bibinfo{author}{Schneider, H.~C.}
\newblock \bibinfo{title}{Simulation of hot-carrier dynamics and terahertz
  emission in laser-excited metallic bilayers}.
\newblock \emph{\bibinfo{journal}{Phys. Rev. Applied}}
  \textbf{\bibinfo{volume}{11}}, \bibinfo{pages}{054083}
  (\bibinfo{year}{2019}).
\newblock
  \urlprefix\url{https://link.aps.org/doi/10.1103/PhysRevApplied.11.054083}.
\newblock \doiprefix 10.1103/PhysRevApplied.11.054083.

\bibitem{Kaltenborn:2014du}
\bibinfo{author}{Kaltenborn, S.} \& \bibinfo{author}{Schneider, H.~C.}
\newblock \bibinfo{title}{{Spin-orbit coupling effects on spin-dependent
  inelastic electronic lifetimes in ferromagnets}}.
\newblock \emph{\bibinfo{journal}{Physical Review B}}
  \textbf{\bibinfo{volume}{90}}, \bibinfo{pages}{201104--5}
  (\bibinfo{year}{2014}).

\bibitem{Zhukov}
\bibinfo{author}{{V. Zhukov, E. Chulkov, and P. Echenique}}.
\newblock \bibinfo{title}{Lifetimes and inelastic mean free path of low-energy
  excited electrons in fe, ni, pt, and au: ab initio gw+ t calculations}.
\newblock \emph{\bibinfo{journal}{Phys. Rev. B}} \textbf{\bibinfo{volume}{73}},
  \bibinfo{pages}{125105} (\bibinfo{year}{2006}).

\bibitem{RenGuanhua:825001}
\bibinfo{author}{Guanhua, R.} \emph{et~al.}
\newblock \bibinfo{title}{Terahertz dielectric properties of single-crystal
  mgo}.
\newblock \emph{\bibinfo{journal}{Infrared and Laser Engineering}}
  \textbf{\bibinfo{volume}{46}}, \bibinfo{pages}{825001}
  (\bibinfo{year}{2017}).

\bibitem{Querry1985}
\bibinfo{author}{{M. R. Querry}}.
\newblock \bibinfo{title}{Optical constants}.
\newblock \emph{\bibinfo{journal}{Contractor Report}}
  \textbf{\bibinfo{volume}{8}}, \bibinfo{pages}{CRDC--CR--85034}
  (\bibinfo{year}{1985}).

\bibitem{PhysRevLett.117.207204}
\bibinfo{author}{Belashchenko, K.~D.}, \bibinfo{author}{Kovalev, A.~A.} \&
  \bibinfo{author}{van Schilfgaarde, M.}
\newblock \bibinfo{title}{Theory of spin loss at metallic interfaces}.
\newblock \emph{\bibinfo{journal}{Phys. Rev. Lett.}}
  \textbf{\bibinfo{volume}{117}}, \bibinfo{pages}{207204}
  (\bibinfo{year}{2016}).
\newblock
  \urlprefix\url{https://link.aps.org/doi/10.1103/PhysRevLett.117.207204}.
\newblock \doiprefix 10.1103/PhysRevLett.117.207204.

\bibitem{PhysRevLett.119.017202}
\bibinfo{author}{Alekhin, A.} \emph{et~al.}
\newblock \bibinfo{title}{Femtosecond spin current pulses generated by the
  nonthermal spin-dependent seebeck effect and interacting with ferromagnets in
  spin valves}.
\newblock \emph{\bibinfo{journal}{Phys. Rev. Lett.}}
  \textbf{\bibinfo{volume}{119}}, \bibinfo{pages}{017202}
  (\bibinfo{year}{2017}).
\newblock
  \urlprefix\url{https://link.aps.org/doi/10.1103/PhysRevLett.119.017202}.
\newblock \doiprefix 10.1103/PhysRevLett.119.017202.

\bibitem{PhysRevLett.112.106602}
\bibinfo{author}{Rojas-S\'anchez, J.-C.} \emph{et~al.}
\newblock \bibinfo{title}{Spin pumping and inverse spin hall effect in
  platinum: The essential role of spin-memory loss at metallic interfaces}.
\newblock \emph{\bibinfo{journal}{Phys. Rev. Lett.}}
  \textbf{\bibinfo{volume}{112}}, \bibinfo{pages}{106602}
  (\bibinfo{year}{2014}).
\newblock
  \urlprefix\url{https://link.aps.org/doi/10.1103/PhysRevLett.112.106602}.
\newblock \doiprefix 10.1103/PhysRevLett.112.106602.

\bibitem{PhysRevLett.105.027203}
\bibinfo{author}{Battiato, M.}, \bibinfo{author}{Carva, K.} \&
  \bibinfo{author}{Oppeneer, P.~M.}
\newblock \bibinfo{title}{Superdiffusive spin transport as a mechanism of
  ultrafast demagnetization}.
\newblock \emph{\bibinfo{journal}{Phys. Rev. Lett.}}
  \textbf{\bibinfo{volume}{105}}, \bibinfo{pages}{027203}
  (\bibinfo{year}{2010}).
\newblock
  \urlprefix\url{https://link.aps.org/doi/10.1103/PhysRevLett.105.027203}.
\newblock \doiprefix 10.1103/PhysRevLett.105.027203.

\bibitem{PhysRevLett.116.196601}
\bibinfo{author}{Battiato, M.} \& \bibinfo{author}{Held, K.}
\newblock \bibinfo{title}{Ultrafast and gigantic spin injection in
  semiconductors}.
\newblock \emph{\bibinfo{journal}{Phys. Rev. Lett.}}
  \textbf{\bibinfo{volume}{116}}, \bibinfo{pages}{196601}
  (\bibinfo{year}{2016}).
\newblock
  \urlprefix\url{https://link.aps.org/doi/10.1103/PhysRevLett.116.196601}.
\newblock \doiprefix 10.1103/PhysRevLett.116.196601.

\bibitem{Penn:1985wt}
\bibinfo{author}{Penn, D.~R.}, \bibinfo{author}{Apell, P.} \&
  \bibinfo{author}{Girvin, S.~M.}
\newblock \bibinfo{title}{{Spin polarization of secondary electrons in
  transition metals: $_{Theory }$}}.
\newblock \emph{\bibinfo{journal}{Physical Review B}}
  \textbf{\bibinfo{volume}{32}}, \bibinfo{pages}{1--16} (\bibinfo{year}{1985}).

\bibitem{Zhu2014}
\bibinfo{author}{Zhu, Y.-H.}, \bibinfo{author}{Xu, D.-H.} \&
  \bibinfo{author}{Geng, A.-C.}
\newblock \bibinfo{title}{Charge accumulation due to spin transport in magnetic
  multilayers}.
\newblock \emph{\bibinfo{journal}{Physica B: Condensed Matter}}
  \textbf{\bibinfo{volume}{446}}, \bibinfo{pages}{43--48}
  (\bibinfo{year}{2014}).

\bibitem{Faghihi:2017vv}
\bibinfo{author}{{Faghihi}, D.} \emph{et~al.}
\newblock \bibinfo{title}{{Moment Preserving Constrained Resampling with
  Applications to Particle-in-Cell Methods}}.
\newblock \emph{\bibinfo{journal}{arXiv e-prints}}
  \bibinfo{pages}{arXiv:1702.05198} (\bibinfo{year}{2017}).
\newblock \eprint{1702.05198}.

\bibitem{PhysRevLett.85.844}
\bibinfo{author}{Koopmans, B.}, \bibinfo{author}{van Kampen, M.},
  \bibinfo{author}{Kohlhepp, J.~T.} \& \bibinfo{author}{de~Jonge, W. J.~M.}
\newblock \bibinfo{title}{Ultrafast magneto-optics in nickel: Magnetism or
  optics?}
\newblock \emph{\bibinfo{journal}{Phys. Rev. Lett.}}
  \textbf{\bibinfo{volume}{85}}, \bibinfo{pages}{844--847}
  (\bibinfo{year}{2000}).
\newblock \urlprefix\url{https://link.aps.org/doi/10.1103/PhysRevLett.85.844}.
\newblock \doiprefix 10.1103/PhysRevLett.85.844.

\bibitem{Werner:2009gt}
\bibinfo{author}{Werner, W. S.~M.}, \bibinfo{author}{Glantschnig, K.} \&
  \bibinfo{author}{Ambrosch-Draxl, C.}
\newblock \bibinfo{title}{{Optical Constants and Inelastic Electron-Scattering
  Data for 17 Elemental Metals}}.
\newblock \emph{\bibinfo{journal}{Journal of Physical and Chemical Reference
  Data}} \textbf{\bibinfo{volume}{38}}, \bibinfo{pages}{1013--1092}
  (\bibinfo{year}{2009}).

\bibitem{stephens1952index}
\bibinfo{author}{Stephens, R.~E.} \& \bibinfo{author}{Malitson, I.~H.}
\newblock \bibinfo{title}{Index of refraction of magnesium oxide}.
\newblock \emph{\bibinfo{journal}{Journal of Research of the National Bureau of
  Standards}} \textbf{\bibinfo{volume}{49}}, \bibinfo{pages}{249--252}
  (\bibinfo{year}{1952}).

\bibitem{Ordal:85}
\bibinfo{author}{Ordal, M.~A.}, \bibinfo{author}{Bell, R.~J.},
  \bibinfo{author}{Alexander, R.~W.}, \bibinfo{author}{Long, L.~L.} \&
  \bibinfo{author}{Querry, M.~R.}
\newblock \bibinfo{title}{Optical properties of fourteen metals in the infrared
  and far infrared: Al, co, cu, au, fe, pb, mo, ni, pd, pt, ag, ti, v, and w.}
\newblock \emph{\bibinfo{journal}{Appl. Opt.}} \textbf{\bibinfo{volume}{24}},
  \bibinfo{pages}{4493--4499} (\bibinfo{year}{1985}).
\newblock \urlprefix\url{http://ao.osa.org/abstract.cfm?URI=ao-24-24-4493}.
\newblock \doiprefix 10.1364/AO.24.004493.

\bibitem{Chandler-Horowitz}
\bibinfo{author}{Chandler-Horowitz, D.} \& \bibinfo{author}{Amirtharaj, P.~M.}
\newblock \bibinfo{title}{High-accuracy, midinfrared (450 cm$^{-1}$ $\leq$
  $\omega$ $\leq$ 4000 cm$^{-1}$) refractive index values of silicon}.
\newblock \emph{\bibinfo{journal}{Journal of Applied Physics}}
  \textbf{\bibinfo{volume}{97}}, \bibinfo{pages}{123526}
  (\bibinfo{year}{2005}).
\newblock \doiprefix 10.1063/1.1923612.

\bibitem{Binder_book}
\bibinfo{author}{Binder, R.}
\newblock \emph{\bibinfo{title}{Optical Properties of Graphene}}
  (\bibinfo{publisher}{World Scientific Publishing Co. Pte. Ltd.},
  \bibinfo{year}{2017}).

\bibitem{Sipe:87}
\bibinfo{author}{Sipe, J.~E.}
\newblock \bibinfo{title}{New green-function formalism for surface optics}.
\newblock \emph{\bibinfo{journal}{J. Opt. Soc. Am. B}}
  \textbf{\bibinfo{volume}{4}}, \bibinfo{pages}{481--489}
  (\bibinfo{year}{1987}).
\newblock \urlprefix\url{http://josab.osa.org/abstract.cfm?URI=josab-4-4-481}.
\newblock \doiprefix 10.1364/JOSAB.4.000481.

\bibitem{Jepsen:96}
\bibinfo{author}{Jepsen, P.~U.}, \bibinfo{author}{Jacobsen, R.~H.} \&
  \bibinfo{author}{Keiding, S.~R.}
\newblock \bibinfo{title}{Generation and detection of terahertz pulses from
  biased semiconductor antennas}.
\newblock \emph{\bibinfo{journal}{J. Opt. Soc. Am. B}}
  \textbf{\bibinfo{volume}{13}}, \bibinfo{pages}{2424--2436}
  (\bibinfo{year}{1996}).

\end{thebibliography}
\section*{Acknowledgments}

We acknowledge the support from the Deutsche Forschungsgemeinschaft (DFG) through the collaborative research center SFB TRR 173: SPIN+X Projects B07, B03. E.~T.~P. acknowledges the Carl-Zeiss-Foundation for the financial support. D.~M.~N.~acknowledges financial support from the Graduate School of Excellence MAINZ (Excellence Initiative DFG/GSC 266). R.~H.~B. thanks TU Kaiserslautern for hospitality and SFB/TRR 173 for support. We thank Prof. Burkard Hillebrands for lab and scientific support.

\subsection*{Author contributions}

E.~T.~P. and R.~B.~conceived the experiments. L.~S. fabricated the samples. G.~T., L.~S.~ and D.~S. carried out the terahertz experiments. J.~L. carried out the XRD measurements. A.~B. performed the TEM structural measurements and the analysis. The experimental THz-data were analyzed by S.~K., G.~T., R.~B.~ L.~S.,  D.~S,  M.~R and E.~T.~P. The theoretical model was developed by D.~M.~N., R.~H.~B., and H.~C.~S.. M.~B.~performed initial calculations. The manuscript was initially written by S.~K., D.~M.~N. and E.~T.~P. All of the authors participated in the discussion, interpreted the results, and reviewed the manuscript.

\section*{Additional information}

\textbf{Competing interests}:
The authors declare no competing interests and non-financial interests.

\end{document}